\documentclass[]{article}
\usepackage{lmodern}
\usepackage{amssymb,amsmath}
\usepackage{ifxetex,ifluatex}
\usepackage{fixltx2e} % provides \textsubscript
\ifnum 0\ifxetex 1\fi\ifluatex 1\fi=0 % if pdftex
  \usepackage[T1]{fontenc}
  \usepackage[utf8]{inputenc}
\else % if luatex or xelatex
  \ifxetex
    \usepackage{mathspec}
  \else
    \usepackage{fontspec}
  \fi
  \defaultfontfeatures{Ligatures=TeX,Scale=MatchLowercase}
\fi
\IfFileExists{upquote.sty}{\usepackage{upquote}}{}
\IfFileExists{microtype.sty}{%
\usepackage{microtype}
\UseMicrotypeSet[protrusion]{basicmath} % disable protrusion for tt fonts
}{}
\usepackage[margin=1in]{geometry}
\usepackage{hyperref}
\hypersetup{unicode=true,
            pdftitle={A Hierarchical Bayes Approach to Adjust for Selection Bias in Before-After Analyses of Vision Zero Policies},
            pdfauthor={Jonathan Auerbach, Christopher Eshleman and Rob Trangucci},
            pdfborder={0 0 0},
            breaklinks=true}
\usepackage{longtable,booktabs}
\usepackage{graphicx,grffile}
\makeatletter
\def\maxwidth{\ifdim\Gin@nat@width>\linewidth\linewidth\else\Gin@nat@width\fi}
\def\maxheight{\ifdim\Gin@nat@height>\textheight\textheight\else\Gin@nat@height\fi}
\makeatother
\setkeys{Gin}{width=\maxwidth,height=\maxheight,keepaspectratio}
\IfFileExists{parskip.sty}{%
\usepackage{parskip}
}{% else
\setlength{\parindent}{0pt}
\setlength{\parskip}{6pt plus 2pt minus 1pt}
}
\ifx\paragraph\undefined\else
\let\oldparagraph\paragraph
\renewcommand{\paragraph}[1]{\oldparagraph{#1}\mbox{}}
\fi
\ifx\subparagraph\undefined\else
\let\oldsubparagraph\subparagraph
\renewcommand{\subparagraph}[1]{\oldsubparagraph{#1}\mbox{}}
\fi

\let\rmarkdownfootnote\footnote%
\def\footnote{\protect\rmarkdownfootnote}

\usepackage{titling}

  \title{A Hierarchical Bayes Approach to Adjust for Selection Bias in
Before-After Analyses of Vision Zero Policies}
  \pretitle{\vspace{\droptitle}\centering\huge}
  \posttitle{\par}
  \author{Jonathan Auerbach, Christopher Eshleman and Rob Trangucci}
  \preauthor{\centering\large\emph}
  \postauthor{\par}
  \predate{\centering\large\emph}
  \postdate{\par}
  \date{August 1, 2018}

\usepackage{bbm}

\begin{document}
\maketitle

\vfill

\subsection{Abstract}\label{abstract}

American cities devote significant resources to the implementation of
traffic safety countermeasures that prevent pedestrian fatalities.
However, the before-after comparisons typically used to evaluate the success of
these countermeasures often suffer from selection bias. This paper motivates
the tendency for selection bias to overestimate the benefits of traffic safety policy, using New York 
City's Vision Zero strategy as an example. The NASS General Estimates 
System, Fatality Analysis Reporting System and other databases are combined 
into a Bayesian hierarchical model to calculate a more realistic before-after comparison. The 
results confirm the before-after analysis of New York City's Vision Zero policy did in fact 
overestimate the effect of the policy, and a more realistic estimate is roughly two-thirds the size.

\newpage

\subsection{1. Introduction}\label{introduction}

Researchers cite motor vehicle collisions as a leading source of
preventable death in the United States (Mokdad et al. 2004). Thirty-five
thousand Americans were killed by motor vehicles in 2013. That exceeds
the number of fatalities by more controversial causes such as firearms
(thirty-four thousand deaths) and alcohol (twenty-nine thousand deaths)
(Xu et al. 2016). Yet gun and alcohol related deaths receive a
significantly larger portion of the public's attention.

The difference between collisions and guns or alcohol is that policymakers
have historically considered collisions unavoidable. Traditional road design 
manuals explicitly recognized a tradeoff between safety and mobility. 
Engineers were instructed to build roads to accommodate the projected traffic 
volume and reduce bottlenecks. For example, urban arterials were designed to 
achieve a traffic flow of 30 to 60 miles per hour (mph) (Aashto 2001). Policymakers 
then set speed limits to the 85th percentile speed of drivers observed in 
favorable operating conditions. This speed was thought to offer a reasonable 
balance between safety and mobility, and drivers were expected to reduce 
their speed only when unfavorable operating conditions produced increased 
risk (National Research Council (US). Transportation Research Board and Limits 1998).

An increasing number of American cities have found the traditional
approach inadequate. Urban drivers are too often found unresponsive to 
the conditions that put more vulnerable road users, like pedestrians, 
at increased risk. In 2014, fifteen percent of motor vehicle fatalities were 
pedestrians, and seventy-eight percent of these occurred in urban areas 
(Administration 2016). Twelve major American cities have since established 
comprehensive strategies that compel drivers to make safer decisions: Chicago, 
San Francisco, New York City, Boston, Los Angeles, Austin, Portland, Seattle, 
San Jose, San Diego, Washington D.C. and Denver. These strategies are 
collectively known as Vision Zero and have a stated goal of creating a road 
system with zero traffic fatalities.

The term Vision Zero is not mere rhetoric. It refers to a specific,
long-term infrastructure investment strategy that rejects the traditional
balance between safety and mobility. Vision Zero prioritizes safety over
mobility by anticipating human error and then slowing vehicles to the
safest feasible travel speed. Success is assured by physics: slower
vehicles collide with less force and are less likely to kill a pedestrian. 
For this reason, citywide road redesign that physically slows vehicles
is the basis of Vision Zero as originally conceived (Tingvall and
Haworth 2000).

Although effective in theory, a complete redesign of roads has proven
challenging to implement and sustain across an entire city. A pragmatic albeit 
less assured approach is to adopt countermeasures that encourage vehicles 
to reduce their speed. For example, cities often install signs, improve lighting 
and visibility of cross walks or direct police officers to prioritize enforcement 
of the traffic law. These countermeasures are effective when drivers behave 
as though the road had been redesigned. Needless to say encouragements 
without commensurate levels of design may not prevent fatalities if drivers have 
little incentive to change their speed in compliance.

Consider the posted speed limit. Vision Zero cities frequently encourage
vehicles to reduce their speed by lowering the limit below the speed for
which the road was originally designed. Such speed limit reductions are
common because they can be implemented immediately and at relatively
little cost, and the reduction ostensibly affects every road across an
entire city (Leaf and Preusser 1999). However, the National Highway
Safety Traffic Administration rates the countermeasure ``reduce and
enforce speed limits'' only three out of five stars for improving safety
because research indicates that actual vehicle speed declines
by only a fraction of the reduced limit, typically 1-2 mph for every 5
mph reduced. It states that effectiveness requires the reduced limit be
met with communications and outreach, enforcement and engineering
changes (Goodwin et al. 2010).

Nevertheless, traffic safety advocates promised American cities reductions
in fatalities when countermeasures are combined as part of a larger Vision 
Zero strategy. Many cited European countries like Sweden, which pioneered 
the Vision Zero movement in the mid-1990s (Government Offices of Sweden 
and Investment Council, n.d.). Sweden took steps such as reducing the posted 
speed limit, separating traffic lanes and erecting barriers. Fatalities fell 28 
percent, from 6 to 4.7 deaths per 100,000 residents (Johansson 2009); 
meanwhile, the United States experienced 11 deaths per 100,000 residents 
in 2013. The advocates reasoned that American cities were capable of preventing 
a considerable number of fatalities by emulating Sweden. The policy question is 
whether pedestrian fatalities have been prevented by the Vision Zero strategies  
of major American cities, and if so, how many.

This paper demonstrates the utility of a hierarchical Bayes approach to 
address this question. The demonstration is arranged as follows. 
Section 2 establishes the major methodological problem that arises in 
evaluating the effect of Vision Zero strategies. The decision of which 
roads to select for the policy produces before-after comparisons that typically 
exaggerate the number of fatalities prevented, and New York City's Vision Zero 
road prioritization strategy provides a real world example.

Section 3 reviews how selection bias is corrected by replacing the outcomes 
in the before period with their estimated expectation. A nonparametric empirical
Bayes approach is mathematically appealing because it places no
restrictions on how the expected number of fatalities varies across roads. 
But an empirical application overcorrects the before-after comparison,
underestimating the effect of New York City's Vision Zero strategy. A more 
realistic estimate requires additional data on road characteristics. Much traffic safety 
data is collected retrospectively, imposing sparsity and multicollinearity among 
the characteristics of the relatively small subset of roads actually observed. A hierarchical 
Bayes approach is proposed as a favorable alternative capable of handling these complications.

In Section 4, a hierarchical Bayes analysis is performed using data from the twelve major American 
cities that have established Vision Zero strategies. The posterior distributions of the model parameters 
are interpreted, and a brief discussion is highlighted. The results are then applied to the New York City 
example introduced in Section 2. It suggests the number of fatalities prevented from New York City's 
strategy is two-thirds the amount implied by the before-after comparison. The paper then concludes 
in Section 5 with some final thoughts about the evaluation of Vision Zero policies.

\subsection{2. Identifying Selection Bias in Before-After Analyses of Vision Zero Policies}\label{identifying-selection-bias-in-before-after-analyses-of-vision-zero-policies}

Before-after comparisons are commonly used to evaluate traffic safety policies. A 
before-after comparison examines an outcome before and after the implementation 
of a policy and attributes any change to the policy. The comparison can be made in 
a number of ways. Popular with traffic safety researchers are modification factors, 
which calculate the ratio of the outcome after the policy to the outcome before. The 
AASHTO Highway Safety Manual uses modification factors to evaluate policy. This 
analysis considers the reduction factor, which calculates the difference between the 
before and after outcomes. Policymakers largely consult this measure.

Selection bias refers to the systematic misrepresentation of a policy that
occurs when atypical units are chosen for the policy. This
misrepresentation arises in a variety of contexts, but it is particularly common in 
before-after comparisons when triage is used to select units in lieu of 
randomization---the standard of experimental research. Since random selection
is infeasible for large-scale traffic safety policies like Vision Zero, 
before-after comparisons misrepresent the number of fatalities that were actually 
prevented as a result of a Vision Zero strategy.

The statement that a Vision Zero strategy prevented fatalities is understood 
to concern the causal effect of the strategy on the roads selected
for the strategy after implementation of the strategy. For the selected roads, the 
policymaker directly observes the number of fatalities after implementation. 
However, the policymaker does not observe the number of fatalities that would have
occurred had those roads not been selected for the strategy. The main assumption of
the before-after comparison is that the before period reflects this
unobserved outcome so that comparison with the after period captures
the effect of the strategy as implemented.

This section demonstrates how the assumption is violated when 
policymakers select roads based on the number of fatalities in the before period. 
The fatalities considered ``prevented'' by the before-after comparison will not replicate 
in future years when the same policy is expanded to include roads with typical levels of 
fatalities in the before period. For this reason, the before-after comparison constitutes a 
tenuous basis for setting policy. Subsection 2.1 illustrates a scenario in which 
selection bias arises. Although perhaps simplistic, the scenario is useful for introducing 
the diagnostic and correction strategy used in the remaining analysis. Subsection 2.2 
provides a real world example.

\subsubsection{2.1 Selection bias results in before-after comparisons that typically overestimate the benefits of traffic safety policy}
\label{selection-bias-results-in-before-after-comparisons-that-typically-overestimate-the-benefits-of-traffic-safety-policy}

In practice, triage leads to the selection of roads with an atypically
large number of fatalities in the before period. This causes before-after
comparisons to overestimate the number of fatalities that would 
subsequently occur had the strategy never been applied. The 
number of fatalities is atypically large because selection favors roads with 
fatalities determined in part by erratic fluctuations that are unlikely to repeat
themselves. It is the transient nature of these fluctuations that result in fewer 
fatalities on the selected roads in subsequent periods. Any decrease between 
the before and after periods is then partially an artifact of the selection process, 
and this artificial portion is often referred to as the regression to the mean or simply the
regression effect (Freedman, Pisani, and Purves 1998).

Among statistical artifacts, the regression effect is particularly
pernicious in that large decreases in fatalities can occur on roads even
when a policy has no effect whatsoever, and it is difficult to assess
the extent to which a decrease is an artifact of the selection process
without knowing the population from which the roads were selected. 
The population affords the context necessary for determining how atypical the 
selected roads are, as illustrated with the following scenario.

Consider the following fictional city in which there are one hundred
roads. Assume that in any year, each road independently experiences one
fatality with a probability of ten percent and no fatalities otherwise.
Policymakers would expect ten of the hundred roads to have fatalities.
Suppose that ten roads did in fact experience a fatality.

If the city selected these ten roads for study, but it did not implement
any actual policy changes, there would be one fatality per road on the
selected roads in the before period. Yet, only one of the selected roads
would be expected to have a fatality in a subsequent after period.
Suppose again that this is the case. The reduction factor on selected
roads would show a ninety percent decline in fatalities, from one
hundred to ten percent, even though no policy changes were made. The
selected roads have ``regressed'' towards the mean.

That selection accounts for this decrease can be demonstrated by
calculating the reduction factor on the remaining ninety, unselected
roads. This will mirror the selected roads by ``regressing'' up
towards the mean since if the selected roads were atypical in consisting
of roads of exaggerated danger, the remaining unselected roads would
consist of roads of exaggerated safety. Indeed, the remaining ninety
roads are expected to increase in fatalities from zero to ten percent.

Readers familiar with causal inference might recognize selection bias as a violation 
of the unconfoundedness assumption, and the ``mirror image'' diagnostic in the previous paragraph as 
estimating the effect on pseudo outcomes (Imbens and Rubin 2015) or control outcomes 
(Rosenbaum 2017). To see this, let $B_s$ ($A_s$) denote the observed number of fatalities on the 
selected roads in the before period (after period). Let $B_u$ ($A_u$) denote the number of fatalities 
that would have been observed on the selected roads in the before period (after period) had they not 
been selected. Then the before-after comparison, $A_s - B_s$, can be decomposed as follows:

$$A_s - B_s = (A_s - A_u) + (A_u - B_u) + (B_u - B_s)$$

The first quantity, $(A_s - A_u)$, is the causal effect of the policy as defined in the beginning
of this section. The before-after comparison represents the causal effect when both baselines $(A_u - B_u)$ and
$(B_u - B_s)$ are zero. The first baseline reflects temporal stability. If zero, the number of fatalities would not 
have changed between periods had the selected roads not been selected. The second baseline reflects selection stability.
If zero, the number of fatalities in the before period would not have been different had the selected roads not been selected. 
Selecting the roads with a large number of fatalities in the before period means $B_s$ is atypically large and 
$B_u$ is atypically small. It follows that $(B_u - B_s) < 0$ and $(A_u - B_u) > 0$.

In the fictional city described above, it is observed that $B_s = 10$ and $A_s = 1$. In practice, $B_u$ and
$A_u$ are not observed. However, it can be surmised that $A_u = A_s = 1$, since the policy did nothing, 
and $B_u =0$, since all roads with fatalities were selected. Therefore, $A_s - B_s = -9$, $A_s - A_u = 0$, $B_u - B_s = -10$
and $A_u - B_u = 1$. This last quantity accounts for the ``mirror image'' diagnostic. All roads are comparable 
in the before period because the fatality rate is assumed to follow the same distribution for all roads. The roads 
not selected for study by the fictional city represent what the selected roads would have looked like had 
they not been selected. 

One way to adjust for selection bias is to replace the before outcomes
with the average outcome across all roads in the before period. In the fictional 
city described above, the average number of fatalities in the before period 
is ten percent. The before-after comparison of ten percent to ten
percent gives the correct answer for both selected and unselected
roads: no fatalities were prevented in this fictional city, and the
observed decline is an artifact of the selection process.

This adjustment requires the expected number of fatalities to be the same across 
selected and unselected roads in the before period. Otherwise the average will 
not adequately reflect the future number of fatalities under the conditions of the before 
period. A model is often used to account for the disparity between the selected and 
unselected roads. In fact, the regression model derives its name from its description of the 
regression effect (Stigler 2016).

\subsubsection{2.2 Before-after comparison overestimates the effect of New York City's Vision Zero Strategy}
\label{a-before-after-comparison-that-overestimates-the-effect-of-new-york-citys-vision-zero-strategy}

The real world example is a before-after comparison used to evaluate New
York City's Vision Zero strategy. In November 2014, New York City's
Vision Zero Committee reviewed the location of traffic fatalities and
serious injuries between the years 2009 and 2013. In 2015, these
locations were prioritized for Vision Zero engineering, enforcement and
education countermeasures. A question facing the Committee was how many
pedestrian fatalities were prevented in 2016 as a result of the changes
on these roads (Taskforce 2017). Throughout this analysis, the term
``fatalities'' refers to pedestrian fatalities.

Using the road shapefile provided by the New York State GIS Program
Office, it is estimated that 44,337 road segments, about thirty-two percent 
of New York City road segments, were selected for prioritization.
These road segments made up roughly seventy percent of pedestrian
fatalities between 2009-2013. The priority locations are visualized for
midtown Manhattan in Figure 1. The blue shaded region is the priority
area, the blue lines are priority corridors and the blue triangles are
priority intersections. Blue dots depict fatalities falling within
priority locations. Red dots depict fatalities falling outside priority
locations. Fatalities are displayed in the first panel of Figure 1 for
the last year of the review period, 2013, and in the second panel of
Figure 1 for the year 2016.

As outlined in Subsection 2.1, a decline in fatalities is expected among
the roads selected for prioritization since the selection criteria
favored locations with fatalities in the before period. In fact, a
sizeable reduction was observed and is displayed in the first panel of
Figure 2. Over the 2009-2013 period, an average of 99 pedestrian
fatalities occurred at priority locations throughout the city each year.
In 2016, there were only 72 fatalities at these locations, a 27 percent
decline. The committee's question could be rephrased as asking what
portion of the observed reduction was deaths prevented by New York
City's policy change and what portion was an artifact of the selection
process.

\begin{figure}[htbp]
\centering
\text{Midtown Manhattan road segments selected for prioritization by New York City Vision Zero}
\includegraphics{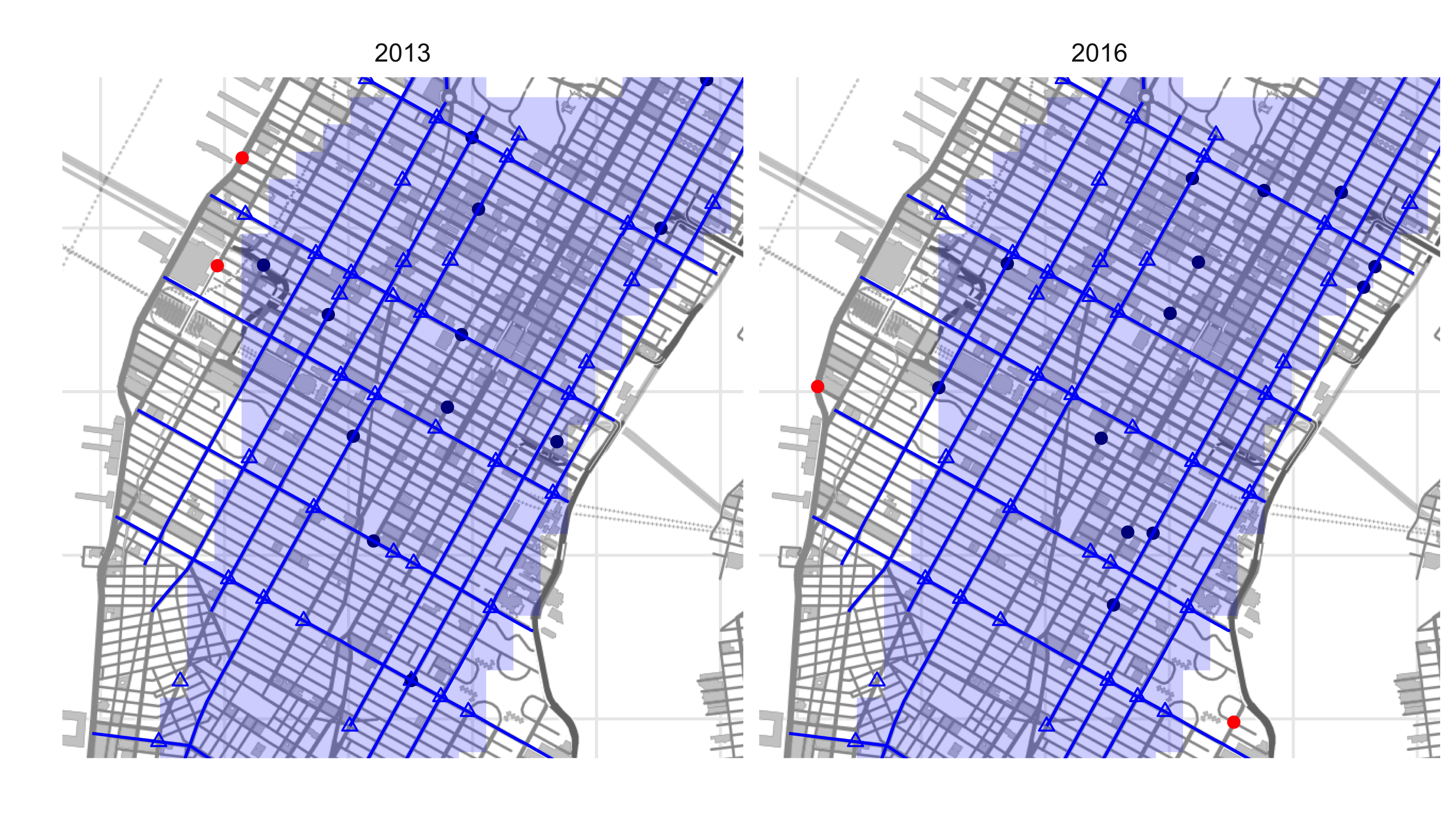}
\caption{This figure displays the locations of pedestrian fatalities in mid-Manhattan for the last year (2013) of the before period (2009-2013) and the after period (2016). Priority corridors are colored blue, priority intersections are triangled and priority zones are shaded blue. Fatalities in priority locations are represented by blue dots, while fatalities outside of priority locations are represented by red dots.}
\end{figure}

Without additional analysis, any division of the observed reduction
between actual prevention and selection bias is possible. On one hand,
the 27 percent decline is nearly equivalent to the reduction experienced
by Sweden after its implementation of Vision Zero. Moreover, a simple
thought experiment using results from the traffic safety literature
explains how this decline might have been achieved in New York City.

Recall, the basis of Vision Zero is the prioritization of safety by
reducing vehicle speed. As part of its strategy, New York City lowered 
the default speed limit from 30 to 25 mph. Suppose for simplicity all 
vehicles traveled uniformly at 30 mph before the City reduced its  
speed limit. A 5mph reduction of the speed limit typically reduces vehicle 
speeds by 1-2 mph so assuming every priority road achieved a 2 mph 
reduction in vehicle speeds, a new speed limit of 25 mph would correspond with
vehicle speeds of 28 mph. According to the regression analysis of Rosén
and Sander (2009), this 2 mph change in vehicle speed would produce a
fatality reduction of 24 percent, three percentage points less than the
reduction suggested by the before-after analysis. Attributing New York City's
27 percent reduction entirely to the Vision Zero Strategy would
therefore be possible if subsequent countermeasures, such as engineering
changes, education and enforcement, led to above average compliance with
the posted speed limit. 

On the other hand, the second panel in Figure 2 suggests a large portion
of the 27 percent reduction in fatalities is driven by selection.
It shows that from the 2009-2013 before period to the 2016 after period,
fatalities at non priority locations increased 20 percent, almost the
same magnitude as fatalities in priority locations decreased. It is
implausible that the strategy caused a substantial number of fatalities
on non priority roads. Instead, the mirror image behavior in Figure 2 is
evidence the procedure used to select priority roads accounts for a
nontrivial portion of the observed reduction in the number of
fatalities.

The portion of the observed reduction due to prevention can be separated
from the portion due to selection by replacing the before values with
their expectations as mentioned in Subsection 2.1. However, the sample
average will not work in this example because the expected number of
fatalities is not the same across selected and unselected roads in the before
period. Dangerous roads were more likely to be selected. A more sophisticated 
approach is necessary to estimate the expectation, and this discussion is the subject 
of Section 3.

\begin{figure}[htbp]
\centering
\text{An increase in fatalities on unselected roads contemporaneous with a decrease on selected road indicates selection bias}
\includegraphics{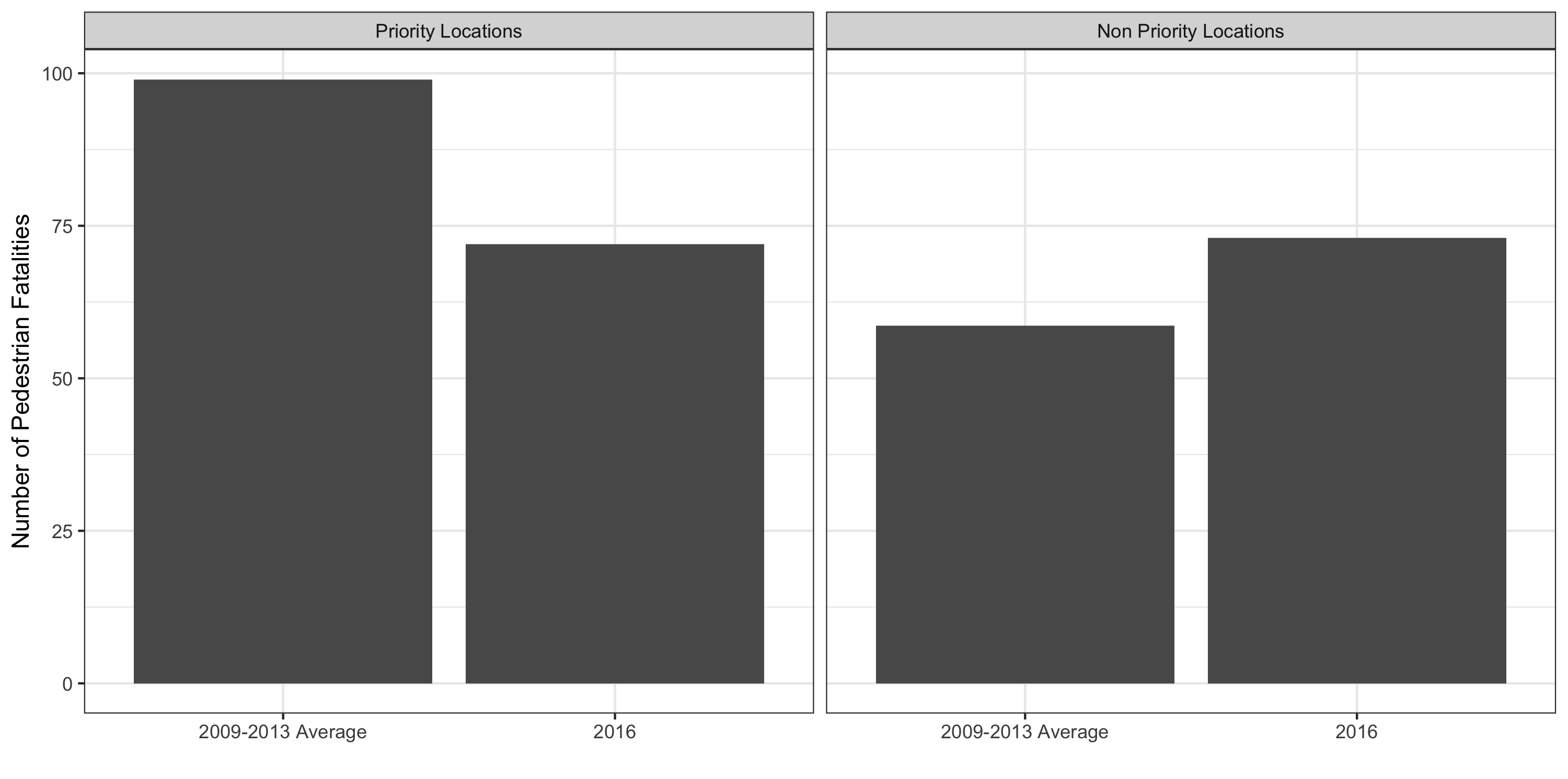}
\caption{This figure displays the number of pedestrian fatalities in New York City in the before and after periods for both priority and non priority roads. The roads selected for priority changes experienced a 27 percent drop in fatalities from an average of 99 over the 2009-2013 time period to 72 in 2016. However, fatalities on the remaining, non priority roads increased 20 percent. This mirror image behavior is evidence of selection bias.}
\end{figure}

\subsection{3. Correcting Selection Bias in Before-After Analyses of Vision Zero Policies}\label{correcting-selection-bias-in-before-after-analyses-of-vision-zero-policies}

Section 2 demonstrates how before-after comparisons exaggerate the benefit of a Vision 
Zero strategy when roads are selected based on the number of fatalities in the before period. 
The before period no longer reflects the number of fatalities one would expect in future periods had 
the selected roads not been selected. The before-after comparison is corrected by replacing the 
number of fatalities in the before period with its expectation, provided no significant change occurs between 
the before and after periods other than the policy. The expectation is estimable by jointly modeling 
the number of fatalities on the selected roads with the number of fatalities on the remaining, unselected roads.

This Section develops a model to estimate this expectation. A nonparametric empirical Bayes analysis 
serves as a mathematically appealing starting point because it places no restrictions on how the expected 
number of fatalities might vary by road. Subsection 3.1 adjusts the before-after comparison using Robbins' 
Formula. However, an application to New York City data underestimates the effect of the Vision 
Zero strategy. A more realistic adjustment is obtained by incorporating additional information on roads. 
Subsection 3.2 presents a hierarchical Bayes regression model that allows for a large number of covariates. 
The model is modified to reflect the retrospective collection of important traffic safety data.

\subsubsection{3.1 A Nonparametric Empirical Bayes Approach Without Informative Covariates Overcorrects for Selection Bias}
\label{a-nonparametric-empirical-bayes-approach-without-informative-covariates-overcorrects-for-selection-bias}

Recognition of the regression effect and selection bias more broadly
lies at the foundation of modern statistical practice, established more than
a century ago (Stigler 2016). Traffic fatalities, along with students' test 
scores and parent/children heights, have long served as the prime examples. 
Despite this history, it remains perhaps the most common pitfall in the statistical 
analysis of data (Friedman 1992). It is no surprise then that multiple corrections 
have been offered. Among the most famous is Herbert Robbins', who 
considered the exact problem of selection bias in before-after studies 
(Robbins and Zhang 1988), and provides a principled starting point. 

The outcome in the before period is described by the following hierarchical 
model. See Gelman and Hill (2006) for an introduction to hierarchical models 
and the accompanying notation:

\begin{align*} 
X_i       &\sim \text{Poisson} (\mu_i) \\
\mu_i &\sim \text{g} (\theta) 
\end{align*}

In the context of a Vision Zero strategy, \(X_i\) denotes the number of pedestrian fatalities 
that might occur on road \(i\) in the before period, $i \in \{ 1, \ldots, I \}$. The rate parameter of each road, 
\(\mu_i\), is thought exchangeable and modeled i.i.d. \(g(\theta)\). The hierarchy is interpreted as describing 
how roads are constructed from a super population of roads that a city could build. From this perspective, the 
Bayes moniker is somewhat of a misnomer. The interpretation is closer to the random effects 
models of classical statistics than a full Bayes analysis. See Berger (2013) for a discussion of this distinction.

Let $I_s < I$ be the number of selected roads. After observing the number of fatalities in the before period, the number of 
fatalities expected on these road had they not been selected is \(E (\sum_{i = 1}^{I_s} \mu_i | X_i) = \sum_{i = 1}^{I_s} E (\mu_i | X_i)\).
Each \(E (\mu_i | X_i)\) is estimable under the above model using Robbins' Formula (Efron and Hastie 2016):

\[\hat{E} ( \mu_i | X_i = x) = (x + 1) \frac{\sum_{i' = 1}^ I \mathbbm{1}(X_{i'}= x + 1)}{\sum_{i' = 1}^I \mathbbm{1}(X_{i'}= x)}\]

Note that index \(i'\) ranges over all roads, $I$, both selected and not selected. This estimate is considered to be a more 
realistic characterization of $\mu_i$ than the road-specific mean, $X_i$. 

The Formula has a long history. Alan Turing originally proposed it 
as a solution to the related Missing Species Problem (Good 1953), but it was 
named after Herbert Robbins, who studied its properties extensively (Robbins 1955,
Robbins and Zhang 2000). Its appeal comes primarily from the fact that
no restrictions are placed on the population distribution, \(g(\theta)\). That is, 
the estimate is derived without any restriction on how the expected number 
of fatalities might vary across each road.

Table 1 applies Robbins' formula to the New York City data from Subsection 2.2. The first row,
\textless{}1\textgreater{} displays the total number of road segments in New York City by the 
number of fatalities between 2009 and 2013. For example, there were 138,142 segments with 
no fatalities, 632 with one fatality and so on. These were plugged into Robbins' Formula to get
\textless{}2\textgreater{}, the expected number of fatalities on each road segment had it not been
selected, having observed the number of fatalities in the before period. This was then multiplied by
\textless{}3\textgreater{}, the number of road segments selected by the Vision Zero Committee as 
priority locations, yielding \textless{}4\textgreater{}, the number of fatalities on priority road segments 
expected over a five year period if not selected. Summing and dividing by the number of years in the 
2009-2013 period gives an average of 52 deaths. Thus despite observing an average of 99 deaths per 
year in priority locations over the 2009-2013 period, adjustment with Robbins' Formula suggests only 
52 fatalities are to be expected in the following year absent any policy change.

Note that the estimated fatality rate is lower for road segments with two fatalities than roads with one fatality. 
This is a known consequence of the Formula (Efron and Hastie 2016). While there are several possible 
corrections, road segments with two or more fatalities are sufficiently rare that adjustments to the estimated 
fatality rate do not significantly change the end result.

The expectation of 52 fatalities per year at the priority locations is 
substantially lower than the 72 fatalities actually observed in 2016
after the implementation of Vision Zero. If the adjustment is to be 
believed, more fatalities have occurred at priority locations than would 
be expected had the previous policy been maintained, not less, as 
suggested by the before-after comparison. Moreover, the increase is 
considerable. The percent increase in the number of fatalities as a result 
of the policy is estimated to be 38 percent.

Yet this 38 percent increase is potentially more pessimistic than a 27
percent decrease is optimistic. Nothing in the traffic safety literature
implies that the countermeasures deployed would substantially increase
the number of fatalities. While the before-after comparison failed to account 
for the fact that many non priority road segments were eligible for selection, Robbins' Formula 
overcorrects by too readily comparing fatalities across priority and non priority road segments.

Additional criteria factored into the selection process, and a portion of the non 
priority road segments have characteristics that make them unlikely to experience a 
fatality and be selected.  For example, Figure 1 suggests that road segments on the 
periphery of New York City were not prioritized, and perhaps the fatalities on these road segments 
should not be compared directly with the priority road segments that were selected. Doing so might
make the priority road segments appear more atypical in the before period than they really are.

Generally speaking, road characteristics, such as the number of lanes or the existence 
of a protective median, inform a Vision Zero selection process and influence the fatality rate. 
These characteristics will be distributed disproportionately across selected and unselected roads. 
Inclusion of this information into a model ensures that the fatalities on the large, four lane roads with 
a protective median likely to be selected are not compared directly to the fatalities on the small one 
way roads unlikely to be selected. The better comparisons are made in the before period, the closer the 
estimated fatality rate in the before period represents what would have happened had a road not been 
selected, and the more accurate the estimated effect of the Vision Zero strategy.

Fortunately, a vast quantity of traffic safety data is available to identify similar roads. This data 
can be used to stratify roads into different types so that a more realistic estimate of the expected 
number of fatalities can be made for each road type. Subsection 3.2 discusses the 
general, methodological complications that arise when incorporating traffic safety data into the 
approach of this section. Section 4 considers data-specific complications.

\begin{longtable}[]{@{}lllll@{}}
\caption{Application of Robbins' Formula to the New York City Vision
Zero data}\tabularnewline
\toprule
Observed Number of Fatalities & 0 & 1 & 2 & 3\tabularnewline
\midrule
\endfirsthead
\toprule
Observed Number of Fatalities & 0 & 1 & 2 & 3\tabularnewline
\midrule
\endhead
\textless{}1\textgreater{} Number of Road Segments & 138,142 & 632 & 40
& 1\tabularnewline
\textless{}2\textgreater{} Estimated Fatality Rate & .0045 & .1266 &
.075 & 4\tabularnewline
\textless{}3\textgreater{} Number of Priority Segments & 43,806 & 405 &
29 & 1\tabularnewline
\textless{}4\textgreater{} Expected Number of Fatalities & 200 & 51 & 
2 & 4\tabularnewline
\bottomrule
\end{longtable}

\subsubsection{3.2 A Hierarchical Bayes Approach Allows for Covariates Informative of Selection}
\label{a-hierarchical-bayes-approach-allows-for-covariates-informative-of-selection}

The empirical Bayes analysis performed in Subsection 3.1 could be salvaged were the selection process 
for a Vision Zero strategy known to depend on a few observable factors. The data could be 
stratified by the factors, and an estimate could be made for the roads in each stratum. The remaining
determinants of fatalities would be evenly distributed among roads within strata, allowing for a fair
comparison of selected and unselected roads. Unfortunately, the selection process depends on a large 
number of factors and, while several databases are available to measure them, doing so introduces 
additional complications that preclude an empirical Bayes approach.
 
A hierarchical Bayes analysis can be viewed as an approximation of the nonparametric empirical 
Bayes approach from Subsection 3.1 that allows covariate information to inform the population distribution, 
\(g(\theta)\). Hierarchical Bayes compares favorably with nonparametric empirical Bayes, even without 
covariates. See Efron and Hastie (2016) and Berger (2013) for a comparison of the methods in general 
and Robbins' Formula specifically.

Suppose traffic safety experts classified each of $i \in \{ 1, \ldots, I \}$ roads into one of \(J \) types. Let 
$j[i]$ denote the type of the $i$th road, $j[i] \in \{ 1, \ldots, J \}$. A univariate hierarchical Bayes log-linear 
regression model describing fatalities in the before period can be written:

\begin{align*} 
X_{i}           &\sim \text{Poisson} ( \mu_i )  \\
           \mu_i &= \exp ( \lambda_{j[i]}) \\
\lambda_j   &\sim \text{Normal} (\theta, \sigma)
\end{align*}

As in Subsection 3.1, \(X_{i}\) denotes the number of fatalities on road \(i\) in the before period. 
However, the log-rate parameter for each road type, \(\lambda_j\), is thought exchangeable and modeled i.i.d. 
$\text{Normal} (\theta, \sigma )$. The hierarchy is still interpreted as describing how roads are 
themselves chosen from a super population of roads that a city could build. 

If the normal distribution were replaced with the arbitrary distribution, \(g(\theta)\), Robbins' Formula 
could be calculated within covariate stratum as in Subsection 3.1. In fact, were experts able to easily classify all roads 
into one of \(J \) types, a number of empirical Bayes approaches would be possible. The negative binomial model is 
common in the traffic safety literature (Hauer 2005), where, in its simplest case, fatalities are assumed to follow a 
Gamma mixture of Poissons, and the hyperparameters are estimated from the data. In fact, the original application of 
the negative binomial was to model accident statistics (Greenwood and Yule 1920).

However, it will be the case that road types are themselves divisible into a large number of subtypes, and roads 
sharing multiple subtypes are more similar than roads sharing fewer subtypes. The univariate hierarchical Bayes 
log-linear regression model is easily extended to this multivariate case, although the notation established in this 
subsection must be augmented. Suppose traffic safety experts classified each road, $i$, into one of 
$j_k \in \{1, \ldots, J_k \}$ subtypes for $k \in \{1, \ldots, K \}$ groups. For example, the first group could describe 
physical road characteristics (e.g. the number of lanes: one lane, two lane, etc.), the second group could describe 
traffic patterns (e.g. the average volume per day: ten vehicles, a hundred vehicles, etc.), the third group could 
describe visibility (e.g. the amount of street lighting: no street lighting, street lighting, etc.), and so on. 
Two roads, $i_1$ and $i_2$, are of the same type, $j \in \{ 1, \ldots, J \}$, if $j_k[i_1] = j_k[i_2]$ for all $k$. 
The multivariate case can now be written:

\begin{align} 
X_{i}           &\sim \text{Poisson} ( \mu_i )  \label{likelihood} \\
           \mu_i &= \exp (\theta + \sum_{k=1}^K \alpha^k_{j_k[i]}) \nonumber \\
\alpha^k_{j_k}   &\sim \text{Normal} (0, \sigma_k) \nonumber
\end{align}

The log-rate parameter for the $i$th road is now the sum of $K$ subtype parameters, $\alpha^1_{j_1[i]} +, \ldots, + \alpha^K_{j_K[i]}$, and the $k$th 
subtype parameter, $\alpha^k_{j_k}$, is thought exchangeable within group $k$ and modeled i.i.d. $\text{Normal} (\theta_k, \sigma_k )$. Note the $\theta_k$ are not identified,
and only the sum, $\theta =  \sum_k \theta_k$, is included in the model above.

In the typical traffic safety study, the covariates that describe each subtype are collected for all roads regardless of the outcome, 
or control roads are carefully chosen for comparison. Neither designs are possible when evaluating Vision Zero due to the 
immense scope of these strategies. Prospective samples are available for only a minority of roads, and they are 
unlikely to contain enough fatalities to describe the potentially complex relationships between the subtype groups and the 
fatality rate. These relationships are better studied using retrospective data, collected in the event of a fatality on a road. 
Estimates obtained from retrospective data can then be weighted back to a Vision Zero city using a prospective sample.

The notation is augmented again to reflect these circumstances. Let \(X_i\) denote the number of fatalities on road 
$i \in \{1, \ldots, I' \}$. Suppose the roads are arranged in decreasing order of fatalities so that the 
first $I < I'$ roads had at least one fatality in the before period and traffic safety experts are only able to classify these roads into one of 
$j_k \in \{1, \ldots, J_k \}$ subtypes for each of the $k \in \{1, \ldots, K \}$ groups.

The typical number of fatalities on the first $I$ roads, $\mu_i$ for $i \leq I$, is estimable from the hierarchical model 
using the zero-truncated Poisson likelihood, $\text{Poisson}^+(\mu_i) =  \frac{\mu_i^x}{ (\exp{(\mu_i)} - 1) x!}$, so 
that equation (\ref{likelihood}) simply becomes:

$$X_{i} \sim \text{Poisson}^+ ( \mu_i )$$

Values or distributions can be chosen for the hyperparameters, and statistical software can simulate \(S \) samples from the 
posterior distribution of the parameters. After model validation, $\mu_i$ can be characterized by the quantiles of $\hat \mu_i(s)$ 
or by $\frac{1}{S} \sum_{s=1}^S \hat \mu_i(s)$, where $\hat \mu_i(s)$ is the $s$th simulation of the posterior distribution of $\mu_i$.
The former corresponds to uncertainty intervals, and the latter corresponds to the posterior mean used in Subsection 3.1.

However, $\mu_i$ cannot be reliably estimated reliably for the remaining $I' - I$ roads with no fatalities and therefore unobserved subtypes. 
These roads cannot be ignored. While the first $I$ roads likely constitute the most dangerous roads in a city and the remaining $I' - I$ roads 
likely have low fatality rates, $I' - I$ will be much larger than $I$ for most cities. It is possible the latter contributes more to the expected 
number of fatalities in the before period. That is, $\sum_{i = I + 1}^{I'} \mu_i  \gg \sum_{i = 1}^I \mu_i$. 

The summand of $\sum_{i = 1}^I \mu_i$ is reweighted to represent the population of roads before the Vision Zero 
strategy.  Let $\mathbbm{1}(i \leq I)$ denote whether the $i$th road was observed in the retrospective data. 
Note that $\sum_{i = 1}^I \mu_i = \sum_{i=1}^{I'} \mathbbm{1}(i \leq I) \mu_i $ and $E(\mathbbm{1}(i \leq I)) = P(X_i > 0)$, 
where $E$ is taken over hypothetical replications of the before period. Inverse probability weights, $\frac{1}{P(X_i > 0)}$, 
can be estimated from a prospective sample to ensure the weighted sum, $\sum_{i = 1}^I \frac{\mu_i}{P(X_i > 0)}$, is 
representative of all roads in a city:

$$E \sum_{i =1}^{I} \frac{\mu_i}{P(X_i > 0)} 
 = E \sum_{i =1}^{I'} \frac{\mu_i \mathbbm{1}(i \leq I)}{P(X_i > 0)} 
 = \sum_{i =1}^{I'} \frac{\mu_i E (\mathbbm{1}(i \leq I))}{P(X_i > 0)} 
 = \sum_{i =1}^{I'} \frac{\mu_i P(X_i > 0)}{P(X_i > 0)} 
 = \sum_{i =1}^{I'} \mu_i$$

The combination of hierarchical Bayes and reweighting is common in the model-based inference of surveys. For example, see 
multilevel regression and poststratification (Gelman and Little 1997, Gelman and Hill 2006, Si, Pillai and Gelman 2015). The 
retrospective data can be viewed as a convenience survey that oversamples roads with covariates predictive of fatalities. 
The prospective sample readjusts the estimates to reflect the desired population of roads. For evaluating Vision Zero strategies, 
the population of interest is the set of selected roads in a city.

\subsection{4. Fitting the Hierarchical Bayes Model to Traffic Safety Data from the Before Period}
\label{fitting-the-hierarchical-bayes-model-to-traffic-safety-data-from-the-before-period}

Section 3 motivates a general hierarchical Bayes model that describes fatalities using covariates informative of selection into a 
Vision Zero strategy. The model is a compromise between the before-after comparison, which incorrectly ignores the fatalities 
on the unselected roads, and the nonparametric empirical Bayes adjustment, which incorrectly compares the fatalities on the 
unselected roads directly with the selected roads. The hierarchical Bayes model provides the correct adjustment if it properly accounts for 
the covariates informative of selection. The remaining determinants of fatalities would balance across the selected and 
unselected roads, enabling a fair comparison and a representative estimate of the fatality rate in the before period.

Yet it is not clear how to properly account for the covariates informative of selection from the data alone. The general hierarchical Bayes model 
described in Subsection 3.2 allows outside information to determine how information is pooled across roads, while the strength of the pooling 
is determined by the data. This Section highlights the specific considerations that arise when fitting the model to police reports from the twelve major American cities 
that have declared Vision Zero policies. The model could be fit with data from a variety of other sources, and the additional complications 
that may arise are discussed in Section 5.

The data presented in this subsection is obtained primarily from two National Highway Traffic Safety Administration sources: the 
\href{ftp://ftp.nhtsa.dot.gov/GES}{National Automotive Sampling System (NASS) General Estimates System (GES)} 
and the \href{ftp://ftp.nhtsa.dot.gov/fars/}{Fatality Analysis Reporting System (FARS)}. GES is a nationally 
representative probability sample selected from the more than five million police-reported crashes that occur annually. 
It is collected prospectively and measures a large number of covariates detailing the road, vehicle and persons involved. The
prospective nature of the dataset is useful for drawing conclusions representative of the unobserved population. The drawback 
is that the sample size is too small to reliably estimate the fatality rate for each road type. Key geographic information is also missing.    

FARS is a retrospective census of fatalities containing a large number of covariates in common with GES. Unlike GES, however, FARS 
includes geographic coordinates that permit the estimation of the traffic and pedestrian density. Subsections 4.1 describes the combination 
of FARS with additional data using these coordinates. A large amount of information is obtained for each road, producing sparsity and 
multicolinearity among some covariates and requiring additional regularization for interpretability. Subsection 4.2 demonstrates  
model interpretation and presents select findings from the fitted model. ANOVA plots, described in Gelman (2005), provide a graphical 
representation of variable importance. Subsection 4.3 discusses the weighting of the results with the GES and applies them to the real 
world example introduced in Subsection 2.2.

\subsubsection{4.1 The Combination of Multiple Datasets Results in Sparsity and Multicolinearity Among the Covariates}
\label{the-combination-of-multiple-datasets-results-in-sparsity-and-multicolinearity-among-the-predictors}

Several datasets are combined with FARS to create the complete, retrospective dataset used in this analysis. 
Missing New York City speed limit data in FARS is imputed from a
\href{http://www.nyc.gov/html/dot/html/about/datafeeds.shtml}{NYC DOT}
shapefile of posted speed limits. The pedestrian population on each road is estimated 
as recommended by the \href{https://www.census.gov/hhes/commuting/data/calculations.html}{Census
Bureau} using the Census Tract Flows \href{https://www.fhwa.dot.gov/planning/census_issues/ctpp/data_products/2006-2010_tract_flows/}{(TPP)}
and the Census Planning Database \href{http://www.census.gov/research/data/planning_database/}{(PDB)}.
Average annual daily traffic is estimated from the Highway Performance Monitoring System
\href{http://www.fhwa.dot.gov/policyinformation/hpms/shapefiles.cfm}{(HPMS)}. 

The combined dataset contains 2,404 fatalities as determined by the
\href{https://safety.fhwa.dot.gov/hsip/spm/conversion_tbl/pdfs/kabco_ctable_by_state.pdf}{KABCO
scale}. It includes every pedestrian death in the twelve major American
cities with Vision Zero policies between 2010 and 2015 for which there
was no missing covariate information that could be reliably estimated
with additional resources. Missing fatalities are assumed to result from
unintentional oversight and considered missing at random. The year 2009 
is dropped because of compatibility issues between GES and FARS, 
and the year 2015 is withheld for model validation.

The nearly twenty-five hundred observations in the dataset are relatively small
compared to the numerous ways roads can differ. The retrospective dataset contains the 
following eight covariate groups describing the road on which the fatality took 
place: the weather and surface condition (COND), the city (CITY), the 
year (YEAR), the posted speed limit (SLIM), the presence of various signs 
or signals (SIGN), the time and lighting (LGHT), the physical road 
characteristics or built environment (BLTE) and the annual average traffic 
density (TFFC). However, these eight covariates are themselves batches of 
hundreds of subtypes that determine the fatality rate and potentially inform selection 
into a Vision Zero strategy. Table 2 displays the first six observations of the dataset 
where numbers denote the various qualitative subtype categories. One quantitative 
variable, the number of pedestrians exposed to the road (EXPR), is also displayed.

The decision to combine covariates in this way came from conversations
with experts, who indicated that the subtypes in a group represent 
similar amounts of information about a road a priori. For example, CITY 
denotes which of the twelve cities contains the road, and it is
expected the contribution of the fatality rate due to the city
containing the road to vary systematically about the average
city level. Interactions between groups is also informative, but the same
conversations suggested that their importance is second order.

Table 3 shows the number of observations with the most common
subtypes for each covariate grouping. A few groupings possess a 
large number of subtypes. COND, LGHT and BLTE have 25, 36
and 111 respectively, and most of these are well represented. For example, 
New York City (12) had around 800 roads, and many roads are observed 
in weather condition 2 (clear weather, dry road surface). But while the marginal 
counts of each subtype can be large, the number of observations for any specific road type 
is small, where road type refers to the interaction of subtypes as defined in Subsection 3.3.

\begin{longtable}[]{@{}rrrrrrrrr@{}}
\caption{Example Observations of Dataset}\tabularnewline
\toprule
COND & CITY & YEAR & SLIM & SIGN & LGHT & BLTE & TFFC &
EXPR\tabularnewline
\midrule
\endfirsthead
\toprule
COND & CITY & YEAR & SLIM & SIGN & LGHT & BLTE & TFFC &
EXPR\tabularnewline
\midrule
\endhead
2 & 1 & 1 & 6 & 1 & 10 & 4 & 1 & 256.52\tabularnewline
2 & 1 & 1 & 6 & 1 & 29 & 4 & 1 & 382.92\tabularnewline
2 & 1 & 1 & 6 & 13 & 29 & 6 & 1 & 859.79\tabularnewline
2 & 1 & 1 & 6 & 3 & 8 & 6 & 1 & 3117.68\tabularnewline
2 & 1 & 1 & 6 & 1 & 8 & 9 & 1 & 3286.61\tabularnewline
2 & 1 & 1 & 6 & 1 & 14 & 18 & 1 & 193.75\tabularnewline
\bottomrule
\end{longtable}

\begin{longtable}[]{@{}llllllll@{}}
\caption{Summary of Covariate Groups}\tabularnewline
\toprule
COND & CITY & YEAR & SLIM & SIGN & LGHT & BLTE & TFFC\tabularnewline
\midrule
\endfirsthead
\toprule
COND & CITY & YEAR & SLIM & SIGN & LGHT & BLTE & TFFC\tabularnewline
\midrule
\endhead
2 :1708 & 12 :721 & 1:354 & 7 :768 & 1 :1423 & 21 :335 & 6 :530 & 1:
55\tabularnewline
23 : 223 & 8 :505 & 2:356 & 8 :523 & 4 : 397 & 8 :298 & 18 :405 &
2:1332\tabularnewline
9 : 202 & 6 :239 & 3:408 & 6 :351 & 3 : 313 & 34 :223 & 14 :147 & 3:
769\tabularnewline
24 : 54 & 9 :159 & 4:406 & 10 :150 & 7 : 69 & 29 :217 & 7 :114 & 4:
120\tabularnewline
3 : 27 & 3 :110 & 5:369 & 14 :127 & 10 : 23 & 1 :189 & 8 :103 &
NA\tabularnewline
18 : 12 & 10 :101 & 6:383 & 9 :127 & 2 : 18 & 25 :135 & 13 : 93 &
NA\tabularnewline
(Other): 50 & (Other):441 & NA & (Other):230 & (Other): 33 & (Other):879
& (Other):884 & NA\tabularnewline
\bottomrule
\end{longtable}

Low counts within combinations of subtypes can lead to interpretability issues 
if the covariate groups become highly associated. Association of the eight covariate groups 
is summarized using Cramer's V statistic in Figure 3. No two groups, aside 
from city and traffic density, are found to be highly associated. It was decided to 
replace TFFC with the covariate ROUT, indicating the road's major function (e.g. 
whether the road segment is a local road, highway, etc).

To contrast Figure 3, association of three-way interactions between select covariate groups is displayed
in Figure 4. The high associations indicate that interactions must either be included 
judiciously in the model or heavily regularized. Judicious inclusion may better promote 
interpretability. However experts disagree on which variables are most important 
and their exact relationship with each other. It was decided to include all two-way 
interactions in the model, and their effects are penalized more strongly than the marginal 
effects.

\begin{figure}[htbp]
\centering
\text{Weak association between most of the covariate groups}
\includegraphics{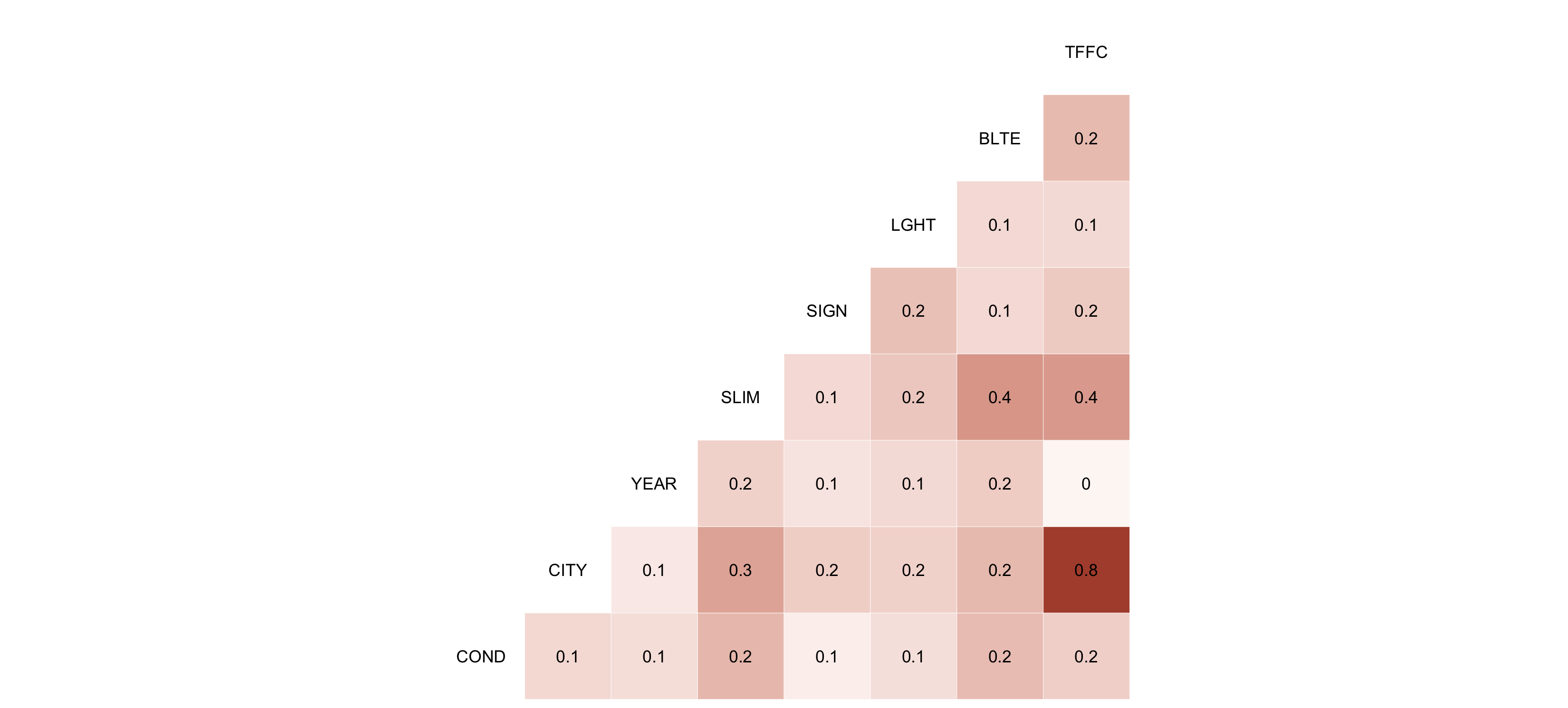}
\caption{This figure exhibits the marginal association of the eight covariate groups using Cramer's V statistic. A value of 0 indicates no association, and a value of 1 indicates complete association. From this figure it is observed that nearly all of the covariates are moderately associated, aside from city and traffic density.}
\end{figure}

\begin{figure}[htbp]
\centering
\text{Strong association between a large number of three-way interactions of covariate groups}
\includegraphics{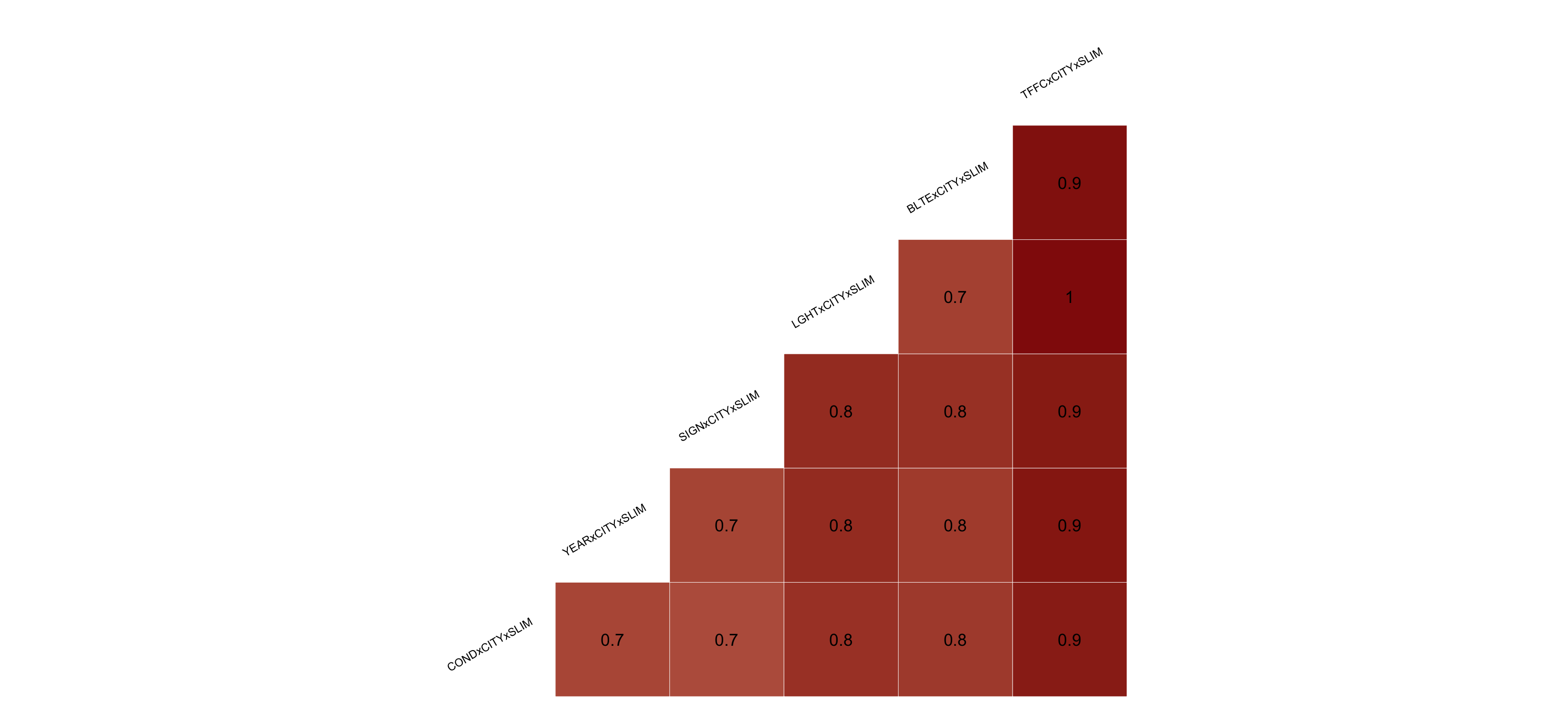}
\caption{This figure exhibits the marginal association of select interactions between covariate groups using Cramer's V statistic. A value of 0 indicates no association, and a value of 1 indicates complete association. From this figure it is observed that all of the interactions are strongly or completely associated.}
\end{figure}

\subsubsection{4.2 Results from the Hierarchical Bayes Approach are Consistent with Previous Vision Zero Findings}
\label{the-hierarchical-bayes-approach-are-consistent-with-previous-vision-zero-findings}

Subsection 4.1 describes the covariates that inform selection into a Vision Zero strategy. These covariates partition the roads 
observed in the before period into $J$ types. Since many factors inform selection, $J$ is large. However, road types with 
common covariates are thought to have similar fatality rates. The covariates are arranged into $K$ groups of subtypes, chosen to 
reflect similar information about each road. 

Let \(X_{i}\) denote the number of fatalities on the $i$th road, and let $j[i]$ denote the type of the $i$th road. As in Subsection 3.2, 
$j[i_1] = j[i_2]$ for any two roads, $i_1$ and $i_2$, if they have all subtypes in common. That is, $j_k[i_1] = j_k[i_2]$ for all $K$ 
groupings.

It is computationally convenient---and without loss of information---to aggregate the number of fatalities across roads of the same type.  
The offset term, EXPR, ensures the aggregates remain comparable. Let $Y_j = \sum_{i:j'[i] = j} X_i$ denote the total number of fatalities 
on all roads of type \(j \), and let $\text{EXPR}_{j}$ denote the total number of pedestrians exposed to all roads of type \(j \). The notation 
$j_k[j]$ refers to the subtype, $j_k$, of the $k$th group for all roads of type $j$. For example, the first group of road subtypes is the speed 
limit, and all roads of the first type have $j_{\text{SLIM}}[j] =  j_1[j] = \text{``road type j has a 30 mph limit''}$. Note 
the $k$th group can be denoted by either a four letter code (e.g. SLIM) or a number (e.g. 1).

When aggregated, the hierarchical Bayes model outlined in Subsection 3.2 describes a high dimensional contingency table and is written:

\begin{align*}
 Y_j &\sim \text{Poisson}^+(\mu_j) \\
 \mu_j &= \text{exp}(\theta + 
 				       \alpha^{\text{SLIM}}_{j_1[j]} +
                                         \alpha^{\text{CITY}}_{j_2[j]} +
                                         \alpha^{\text{YEAR}}_{j_3[j]} +
                                         \alpha^{\text{COND}}_{j_4[j]} \\
& \phantom{= \text{exp}(\theta \  } +
                                         \alpha^{\text{SIGN}}_{j_5[j]} +
                                         \alpha^{\text{LGHT}}_{j_6[j]} +
                                         \alpha^{\text{BLTE}}_{j_7[j]} +
                                         \alpha^{\text{ROUT}}_{j_8[j]} \\
& \phantom{= \text{exp}(\theta \ } +                                         
                                         \beta^{\text{SLIM x CITY}}_{j_9[j]} +
                                         \beta^{\text{SLIM x COND}}_{j_{10}[j]} +
                                         , \ldots, + 
                                         \beta^{\text{BLTE x ROUT}}_{j_{36}[j]} \\
& \phantom{= \text{exp}(\theta \ } +                                                
                                         \epsilon_j +
                                         \rho \cdot
                                         \text{log(} \text{EXPR}_{j})) \\
\alpha^k_{j_k}   &\sim \text{ Normal}(0, \sigma_k) \\
\beta^l_{j_l}   &\sim \text{ Normal}(0, \sigma_l) \\
 \sigma_l &< \sigma_k \\
\epsilon_j     &\sim \text{ Normal}(0, \sigma_{\epsilon}) 
\end{align*}

\vspace{1mm}

All \(\binom 82\) interactions between the covariates in Table 2, excluding EXPR, are included in the model, although
only three are written out above. The standard deviation of the interaction parameters are constrained to be a fraction of the 
standard deviation of the main effects, and as a result the interactions each explain less variation than the marginals. 
The specification of this model is complete with an error term, $\epsilon$ (referred to as ``cell'' error),  and diffuse normal priors are 
placed on \(\mu\), \(\rho\), \(\sigma_k\), \(\sigma_l\) and \(\sigma_{\epsilon}\), as recommended by the 
\href{https://github.com/stan-dev/stan/wiki/Prior-Choice-Recommendations}{Stan Prior Choice Wiki}.

A large number of samples are drawn from the posterior distribution of the parameters given the data, excluding all fatalities in 2015, which
are removed from the dataset for validation. RStan (Stan Development Team 2016) is used to run four chains of two thousand
iterations each. This yields four thousand posterior draws after discarding the first thousand warm-up iterations of each chain. The Stan code 
is provided in the Appendix.

In the Stan generated quantities block, the finite sample standard deviations of each covariate group is calculated (Gelman 2005). 
The number of fatalities of each road type in the dataset in the year 2015 is also predicted for the purpose of checking the validity of the posterior
distribution. An example predictive check is displayed in Figure 5 for New York City. Posterior predictions of the number of
fatalities in the held out test set (the histogram) covers the number actually observed (the dashed line), providing confidence in the 
before-after adjustment computed in Subsection 4.3. Roughly twelve percent of simulated predictions exceed the number of fatalities 
observed in 2015. 

\begin{figure}[htbp]
\centering
\text{Predictions of the number of New York City fatalities in 2015 covers the number actually observed}
\includegraphics{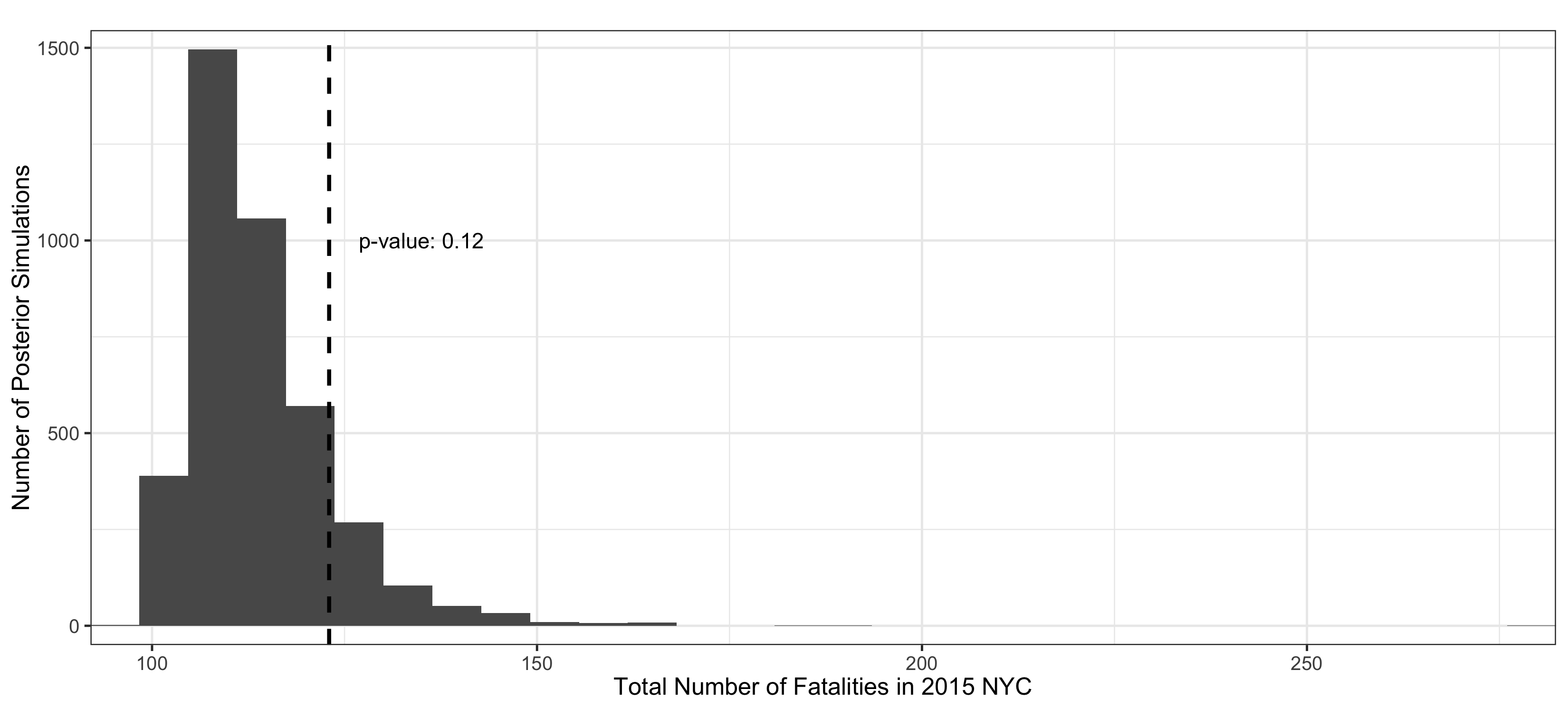}
\caption{This figure exhibits two thousand posterior predictions of the number of pedestrian fatalities in New York City in the held out year 2015 for all road segments that experienced a fatality between 2010 and 2014. The actual number of fatalities observed in 2015 is represented by the dotted line. The proportion of predictions exceeding the number of fatalities observed is twelve percent.}
\end{figure}

Many results are consistent with the experience of traffic safety stakeholders. This consistency also adds confidence in the before-after adjustments. 
Figures 6 displays an Analysis of Variance using the inner 50 and 90 percent intervals of the finite sample standard deviations 
calculated in the generative quantities block. The standard deviations are interpreted as a measure of variable group importance. The fat line represents the 
inner 50 percent range, and the thin line represents the inner 90 percent range. Throughout this analysis, the uncertainty intervals are interpreted as 
corresponding to the likely and plausible locations respectively.

The Analysis of Variance indicates that the city (CITY) and the built environment (BLTE) explain a large proportion of variation in fatalities as expected. However, nearly as 
important is weather and surface condition (COND) and time of day and lighting (LGHT). Conversations revealed these variables capture the varying use of roads each 
day (e.g. commuting, tourism, dinner, nightlife, etc.). There is relatively less variation of the fatality rate within type (cell) and years (YEAR) suggesting that,
through the other covariates, major sources of variation have been accounted for within subtype groups and between the same subtype groups across years.

\begin{figure}[htbp]
\centering
\text{An Analysis of Variance provides a measure of group importance}
\includegraphics{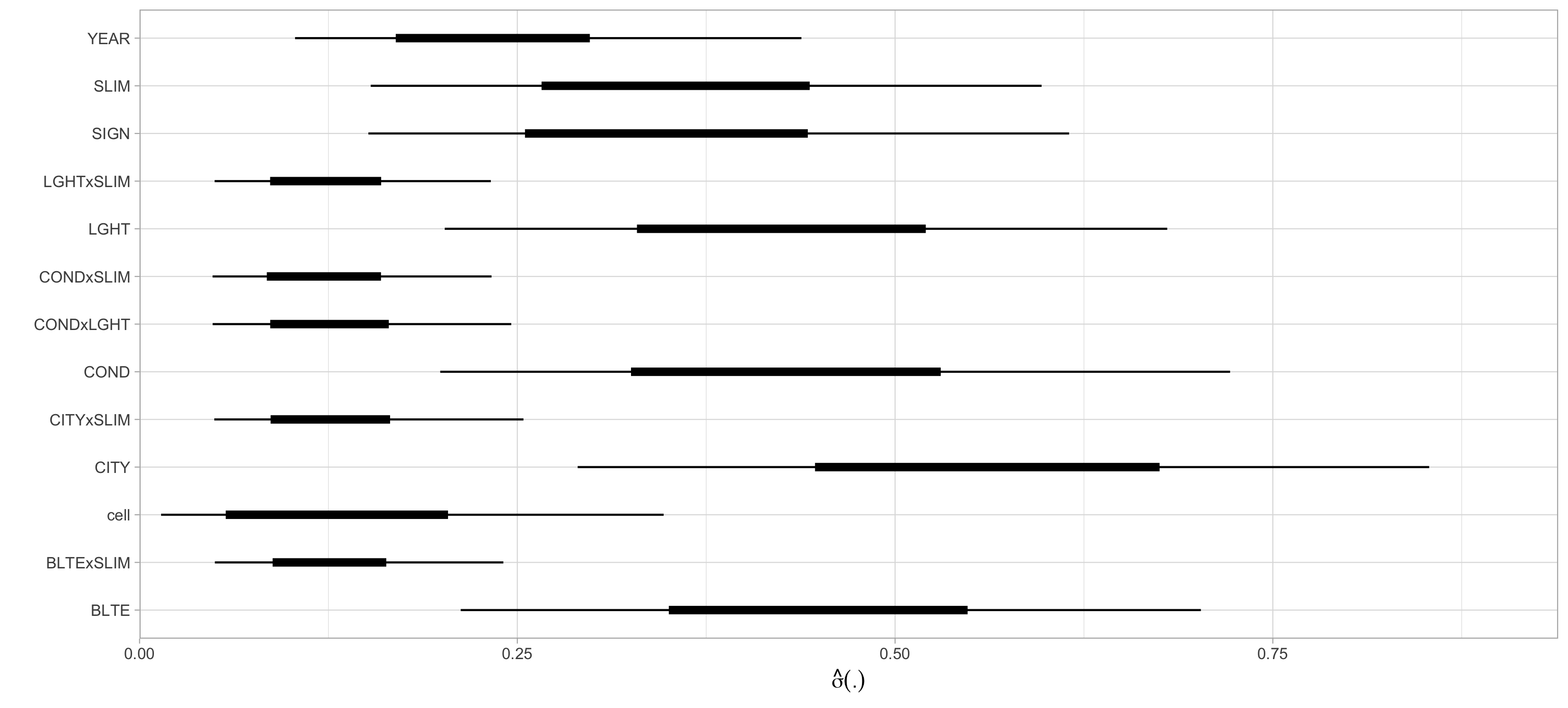}
\caption{In this figure, lines correspond to inner 50 and 90 percent intervals for the finite-population standard deviations of the coefficients of each group of covariates. Select interactions are also included in the Figure.}
\end{figure}

The individual effects were also interpretable. For example, Figure 7 shows the largest LGHT effects, which indicate that the time period of 4:00 pm - 10:00 pm and the 
weekend contain a disproportionate amount of information about fatalities. Conversations revealed that this time period had also been identified by 
researchers as important, for example by the New York City Vision Zero committee (Taskforce 2017).

Other results were suspected by experts but had not been quantified and the magnitude of the effects were surprising. For example, Figure 8 shows that a lot of the 
city variation originates from two cities: Los Angeles and New York City. New York City has an average effect of around 1, which is roughly 2.7 times the average 
city. Meanwhile, Washington D.C. has an average effect of around -.25, which is roughly 78 percent of the average city. Boston was identified in conversation as a 
city with an abnormally large fatality rate. However, Boston does not appear at relatively high risk after adjusting for covariates like the weather and built environment. 

Select interaction effects between CITY and COND are displayed in Figure 9. There is evidence of variation across these categories. For example, weekends from 10 am to 3
pm in New York City and cloudy days in San Diego are disproportionately represented in the data. These interactions could reflect city-specific trends, such as the popularity 
of brunch or the quick advancement of dense fog. While there is too much uncertainty to speculate over the plausibility of these specific explanations, plots of the interactions,
such as in Figure 9, can be used to generate hypotheses that are evaluated in future studies.   

\begin{figure}[htbp]
\centering
\text{Majority of variation in lighting is explained by roads during dawn, dusk or dark and the weekend}
\includegraphics{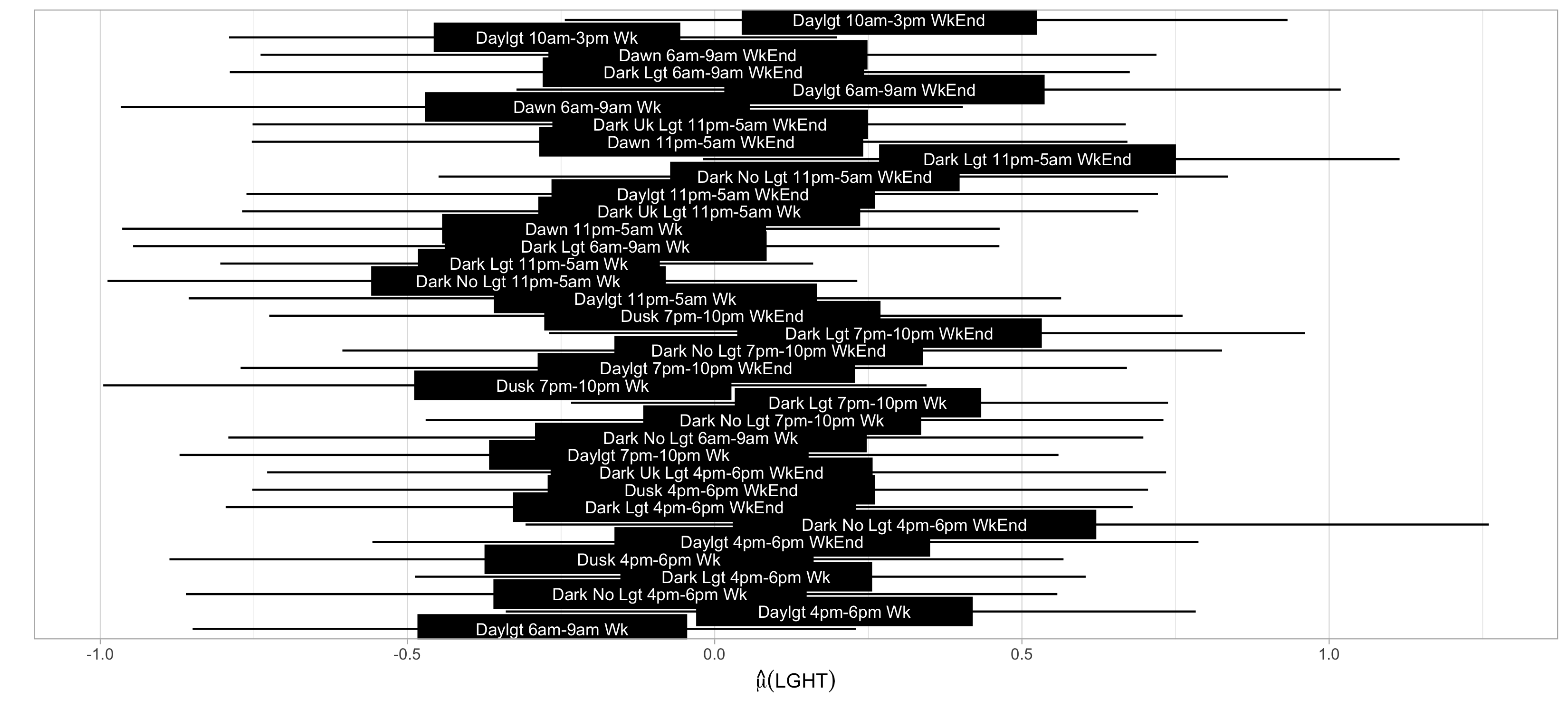}
\caption{This figure exhibits inner 50 and 90 percent intervals for the largest time of day and lighting effects. Effects are interpreted as the log of the expected multiplicative increase in the fatality rate holding all else constant.}
\end{figure}

\begin{figure}[htbp]
\centering
\text{Majority of variation in cities explained by New York and Los Angeles}
\includegraphics{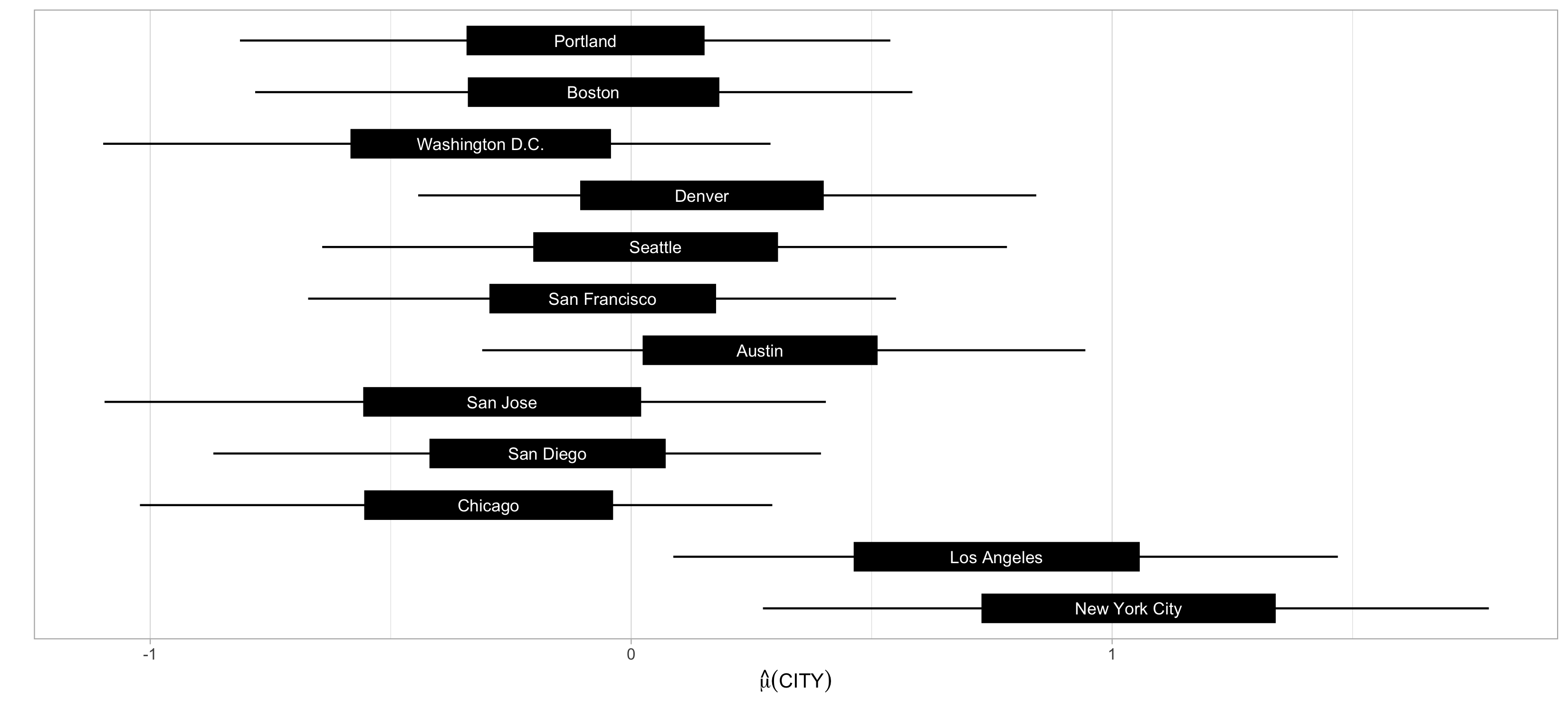}
\caption{This figure exhibits inner 50 and 90 percent intervals for city effects. Effects are interpreted as the log of the expected multiplicative increase in the fatality rate holding all else constant.}
\end{figure}

\begin{figure}[htbp]
\centering
\text{Within city patterns emerge although relatively little variation explained by condition, city interaction}
\includegraphics{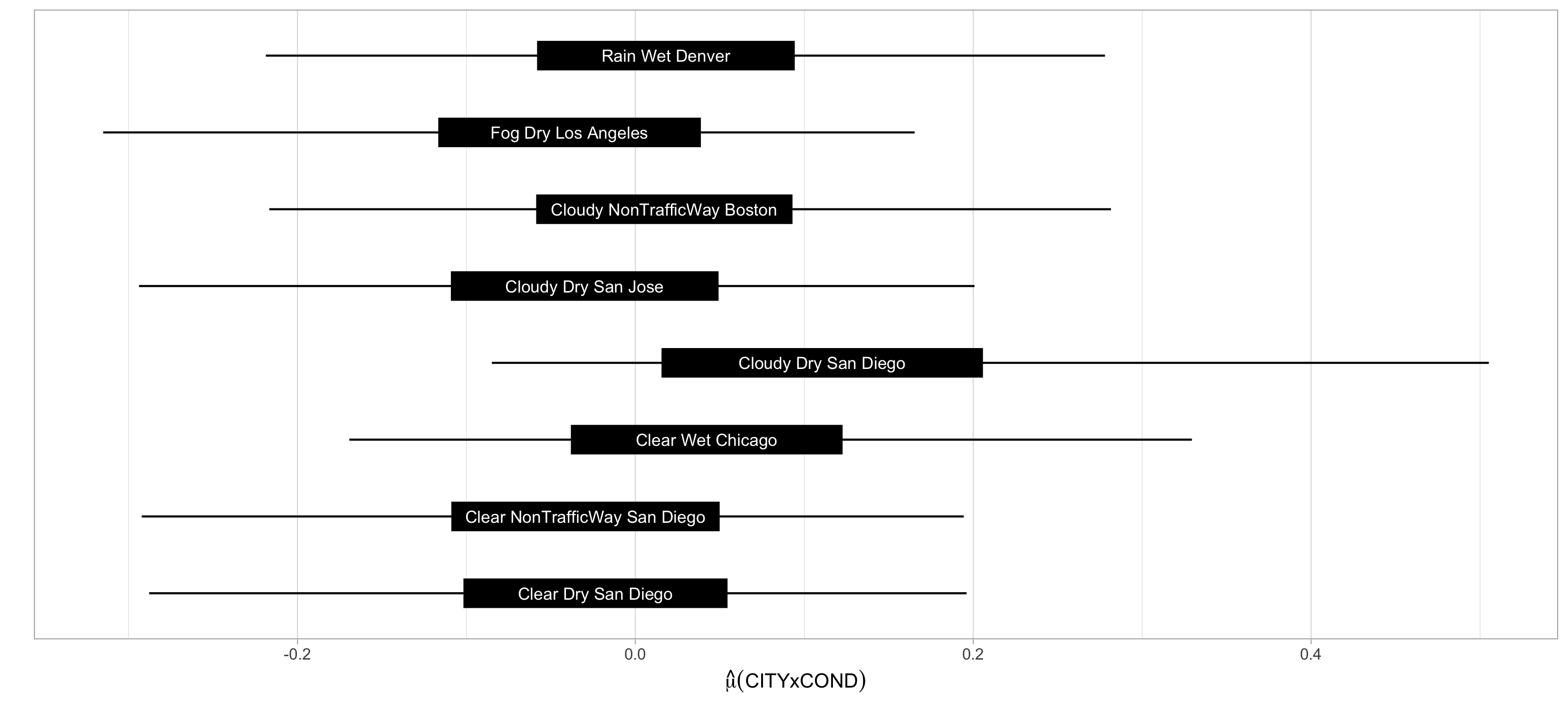}
\caption{This figure exhibits inner 50 and 90 percent intervals for the largest city and weather and surface condition effects. Effects are interpreted as the log of the expected multiplicative increase in the fatality rate holding all else constant.}
\end{figure}

\subsubsection{4.3 The Hierarchical Bayes Approach Provides a Realistic Adjustment for Selection Bias}
\label{the-hierarchical-bayes-approach-provides-a-realistic-adjustment-for-selection-bias}

This subsection concludes the discussion of selection bias in the before-after comparison of New York City's Vision
Zero road prioritization strategy. Sections 2 and 3 present two different estimates for the number of pedestrian fatalities prevented 
as a result of the strategy. A before-after comparison gives a 27 percent decline in fatalities, while adjustment with Robbins' Formula 
gives a 38 percent increase. However, both estimates are unrealistic because selected roads were chosen 
based on the number of fatalities in the before period, and both estimates made unreasonable assumptions about the selection process. 

The before-after comparison assumed the selected roads had a representative number of fatalities in the before period. This was unreasonable 
because New York City deliberately used the number of fatalities to inform the selection process. Conversely, adjustment with Robbins' formula 
assumed all roads were eligible for selection, and this was unreasonable because additional information, such as the function of the road, 
presumably governed its eligibility for selection. Both estimates are extreme, and the actual effect likely resides somewhere in the middle. 
A better estimate could be achieved using additional information governing selection eligibility to compute a different adjustment for each road type.

The importance of determining the type of each road can be seen in Figure 10, which displays the expected number of fatalities for all New York City segment types 
fit with the model described in Subsection 4.2. These densities are not weighted by the number of road segments represented by each type and therefore 
reflect only the New York City road segments on which there was at least one fatality between 2010 and 2013. The density is over point estimates, and does not 
reflect uncertainty in the expected number of fatalities.

A large amount of separation between the road types predictive of fatalities on priority and non priority road segments is observed. The separation
suggests that priority locations do in fact possess covariates more predictive of fatalities than non priority locations and thus have
higher expected fatality rates than estimated by Robbins' formula in Table 1. This explains why adjustment by Robbins' formula gave an
increase in the number of fatalities and not the decrease indicated in the traffic safety literature.

To compute the hierarchical Bayes adjustment for New York City's Vision Zero strategy, samples from the model described in Subsection 4.2 are combined 
with the GES sampling weights to calculate the expected number of fatalities on priority road segments over the 2010-2013 before period as 
described in Subsection 3.2. Let $i \in \{1,\ldots, I \}$ denote a New York City priority road segment observed in the combined, retrospective dataset of 
Subsection 4.1, and $j[i]$ the corresponding road type, $j \in \{1,\ldots, J \}$. Similarly, let $i' \in \{1,\ldots, I' \}$ denote a report in GES on a road of the 
same type so that $j[i] = j[i']$. That is, $\{ i': j[i'] = j[i] \}$ is the set of all GES reports occurring under similar conditions as the fatality on road $i$. As in 
Subsection 3.2, suppose the GES reports are arranged in decreasing order of fatalities so that the first $I'_f < I'$ reports resulted in at least one fatality. The 
probability of a fatality is estimated from the GES as:

$$\hat P(X_i > 0) = \sum_{i': j[i'] = j[i]} \frac{ w^t_{i'} \mathbbm{1}(i' \leq I'_f) + 1}{w^t_{i'} + 1}$$

The $w^t_i$ are computed from the sampling design as follows. Nationally representative weights are constructed in GES by its three stage, stratified 
sampling design: $\frac{1}{w^n_{i'}} = P(\text{PAR} \ | \ \text{PJ})P( \text{PJ} \ | \ \text{PSU} )P( \text{PSU})$, where PAR is the event that a Police Accident 
Report is sampled as observation $i$, PJ is the event that the Police Jurisdiction containing the PAR is sampled and PSU is the event that the 
Primary Sampling Unit containing the PJ is sampled. The target population is a combination of police jurisdictions, and its weight, $w^t_i$, is obtained 
after multiplying the nationally representative weights, $w^n_{i'}$, by the probability a sampled report would be in a PJ in the target population as 
described in the GES documentation (Shelton 1991). Additive smoothing---the addition of 1 to the numerator and denominator---ensures no road is 
assigned a zero probability of a fatality.

Estimates of the expected number of fatalities on priority and non priority road segments, had the Vision Zero strategy never been implemented, are
computed for the three largest Boroughs in New York City: Brooklyn, Manhattan and Queens and displayed in Figure 11. The estimates were computed
using the weights described above in the manner outlined at the end of Subsection 3.2. First, the expected number of fatalities for each road segment observed
in the retrospective data was determined from the fitted model described in Subsection 4.2. Then, the weighted average was taken using the inverse 
probability of inclusion as determined by the above formula. Estimates are only computed for Brooklyn, Manhattan and Queens, and not Staten Island 
and the Bronx, because the latter two Boroughs were not actually sampled in the GES. However, since all boroughs were eligible to be sampled, the 
estimates can be scaled to reflect the entire City.
 
In Figure 11, black lines represent inner fifty and ninety percent intervals for the expected number of fatalities as determined using the posterior simulations
of the fitted model as described in Subsection 4.2. The red lines represent the average number of observed fatalities from 2009-2013 and the blue lines 
represent the number of observed fatalities in 2016. The regression to the mean effect is apparent in the Figure: for both priority and non priority road 
segments, the number of fatalities after Vision Zero moves towards the average number predicted from the model. Notice the blue line for non priority 
roads is located roughly between the black lines, and therefore the model is consistent with the number of fatalities in the after period. By this measure, 
the model accounts for the selection effect. However, the blue line for priority roads is lower than what was expected from the model, and this discrepancy 
is interpreted as reflecting the number of fatalities prevented by the implementation of Vision Zero strategy.

There may be too much uncertainty to quantify the exact number of fatalities prevented by Vision Zero to the satisfaction of policymakers.
The fifty percent uncertainty interval in Figure 16 indicates that there was an average reduction in fatalities between fifteen and thirty-one
percent, and the ninety percent uncertainty interval indicates an average reduction anywhere between one and forty percent. At the
ninety-five percent uncertainty level, it is impossible to completely rule out that the policy prevented no fatalities or that it reduced
fatalities to the maximal attainable reduction suggested in Subsection 2.2.

More nuanced statements can be evaluated using the posterior distribution. For example, the before-after comparison, which suggested
a 27 percent reduction on priority roads in New York City (or for Brooklyn, Manhattan and Queens, a 30 percent decline), underestimates
the actual number of fatalities prevented in twenty-five percent of posterior simulations. That is, after looking at the data, the model gives a 
seventy-five percent chance the expected number of fatalities prevented is as low or lower than the amount suggested by the before-after 
comparison. Conversely, it is estimated that there is a four percent chance that the policy led to no change or an increase in fatalities.

The median reduction for New York City is roughly 18 percent, two-thirds the size of the before-after comparison. This estimate is considered 
a realistic summary of the effect size of the Vision Zero strategy because the median increase on the non priority roads is negligible. 
It is also three-quarters of the 24 percent reduction predicted by Rosén and Sander (2009) and Goodwin et al. (2010) when posted speed 
limits are enforced commensurately.

\begin{figure}[htbp]
\centering
\text{Little overlap in the number of expected fatalities on priority and non priority road types}
\includegraphics{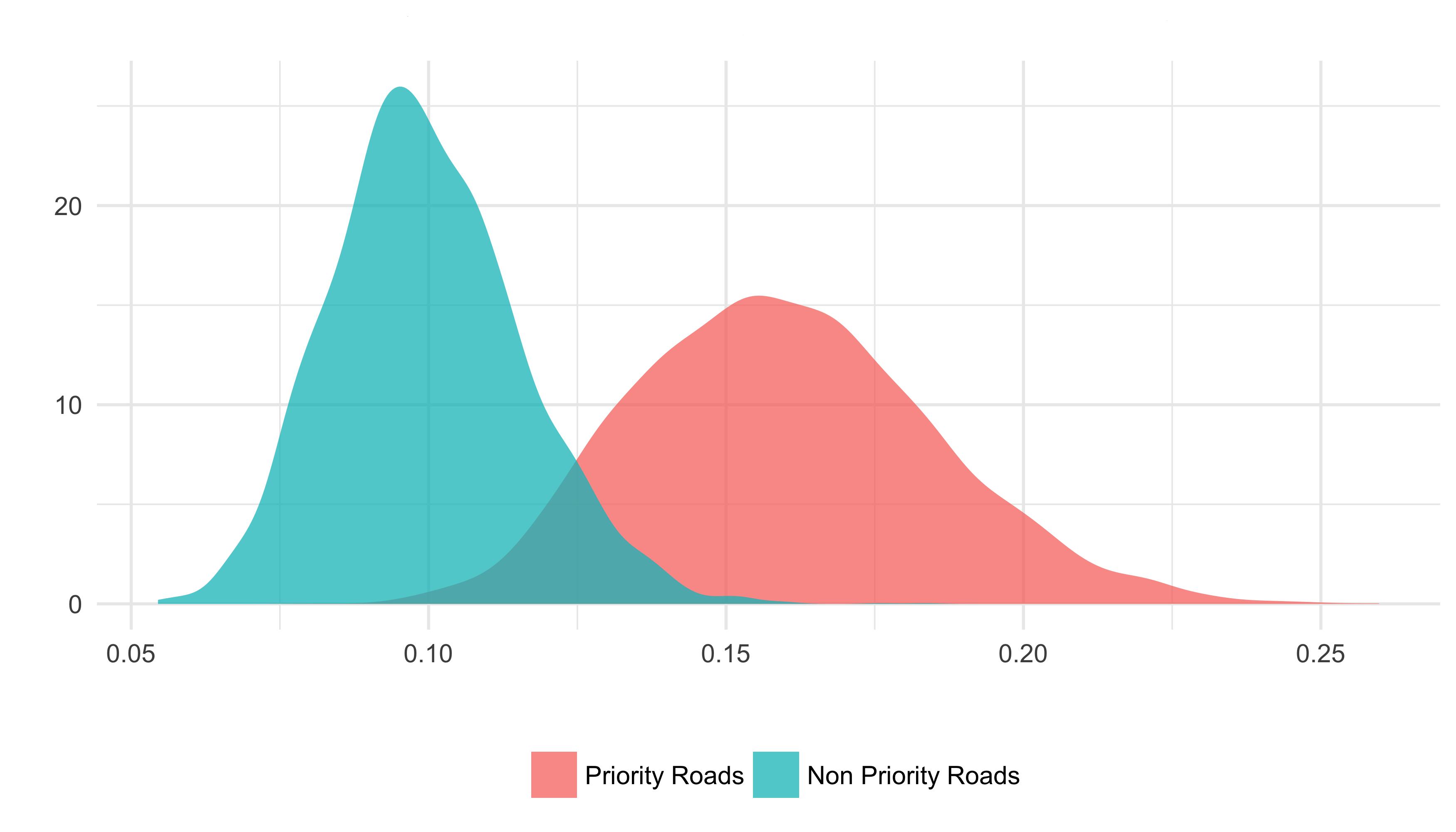}
\caption{This figure exhibits the variation in the estimated expected number of fatalities across different road types in New York City on which there was a fatality between 2010 and 2013. Priority road types have been shaded red while nonpriority road types have been shaded blue.}
\end{figure}

\begin{figure}[htbp]
\centering
\text{Selection bias correction using the proposed model}
\includegraphics{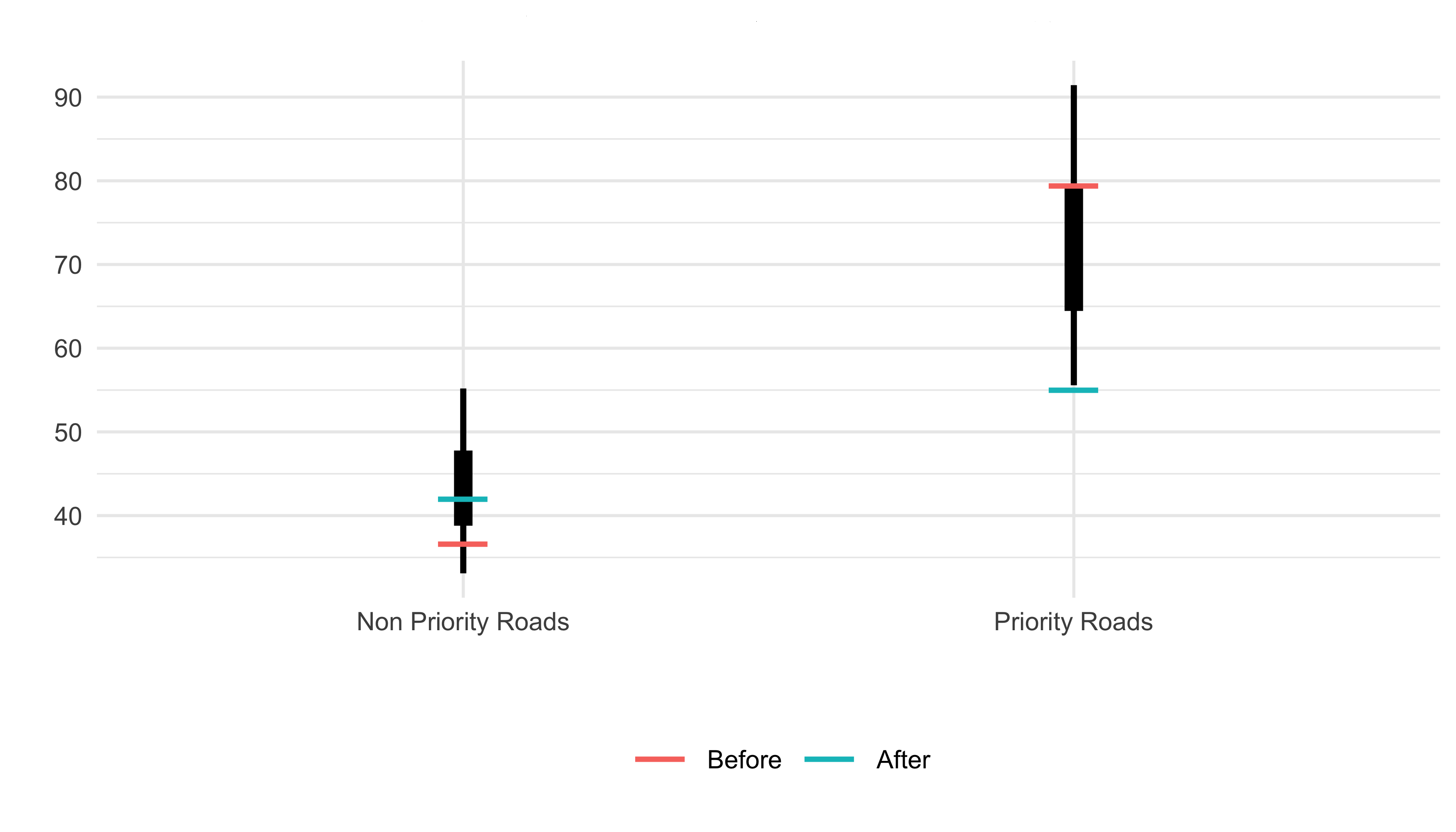}
\caption{This figure exhibits the expected number of fatalities in the New York City Boroughs of Brooklyn, Manhattan and Queens. The red lines mark the average number of fatalities observed on priority and nonpriority road segments over the 2009-2013 before period. The black lines represent 50 and 95 percent uncertainty intervals of the expected number of fatalities in the before period from the Hierarchical Bayes model. The blue lines mark the observed number of fatalities in 2016.}
\end{figure}

\subsection{5. Discussion}\label{discussion}

Over the past decade, major American cities have implemented Vision Zero strategies aimed at preventing pedestrian 
fatalities. The policy question now facing cities is to what extent these strategies have actually prevented fatalities. 
This paper made two main points regarding the before-after comparisons commonly used to answer this question. 

First, it demonstrated how selection bias produces before-after comparisons that exaggerate the effect of a Vision 
Zero strategy. A theoretical explanation of this exaggeration was provided, followed by a real world example 
identifying selection bias in a before-after comparison used to evaluate New York City's Vision Zero road 
prioritization strategy. As predicted by theory, the before-after comparison overestimated the effect anticipated 
by the traffic safety literature and could not explain a massive increase in fatalities on the roads not selected for the 
strategy.

Second, this paper demonstrated how a hierarchical Bayes adjustment provides a more realistic estimate.
This adjustment augmented a simpler nonparametric empirical Bayes analysis, which proved unrealistic when 
applied to the New York City data. Traffic safety experts interpreted the proposed hierarchical Bayes adjustment, 
and when it was applied to the same data, it produced an estimate consistent with the traffic safety literature---and 
roughly two-thirds the size of the unadjusted before-after comparison. It also corrected the statistical 
artifact present in the unadjusted before-after comparison in which a massive increase in fatalities was observed on 
the roads not selected for the strategy. 

This paper was streamlined in order to make these two points most effectively. It deliberately overlooked some potential 
sources of confounding and uncertainty. Yet despite this streamlining, significant effort was still required to identify 
and correct selection bias in a before-after comparison that is popular precisely because it is simple. Moreover, the 
estimate contains a substantial amount of uncertainty.

Some might conclude the effort of adjustment is not worth the reward and that it is easier to avoid selection bias in the first place. 
For example, policymakers might compute a before-after comparison on all roads in a city, regardless of whether they were selected 
for the strategy. While this may avoid selection bias, it does not answer the question to what extent a Vision Zero strategy was effective 
on the selected roads, and it cannot be used to determine whether the strategy should be continued in future years or expanded to
additional roads.

Answering the policy question of interest may mean embracing uncertainty. More accurate measurements, such as better police records, 
might not be feasible, and additional data sources can introduce more uncertainty than they reduce. For example, the exposure 
variable was constructed from the Census tract-level population to represent the number of pedestrians eligible for a fatality. 
Although admittedly broad, a more accurate count of pedestrians on the street may not provide a better measure of the true 
exposure if dangerous behavior, such as j-walking or being distracted, is not evenly distributed among the pedestrian population. Furthermore, 
a count from observational or administrative data may introduce confounding if the collection process differs substantially over space or time.

Future work could better quantify uncertainty in the proposed adjustment. This analysis may have understated uncertainty because the 
covariates and probability weights described in Sections 3 and 4 were treated as fixed constants. However, traffic safety data is prone to 
measurement error and missingness may be informative. Jointly modeling these quantities in the hierarchical Bayes model would produce 
an even more realistic adjustment. In addition to modeling the probability weights, a more robust weighting strategy could be employed. 
Si, Pillai and Gelman (2015) discuss the advantages of Bayesian nonparametric weights in comparison to the inverse probability weights 
used in this analysis.

The primary concern of this paper was to compute a point estimate for the number of fatalities prevented as a result of implementing a Vision Zero strategy, 
including both the countermeasures deployed and all subsequent changes that typically occur in cities. No attempt was made to separate a countermeasures' 
direct impact from its context or execution. It did not adjust for the actual change in traffic conditions, human behavior or other intervening factors. 

This point is worth stressing due to the differing positions on what is meant by a  ``causal effect''. For example, taken literally, the effect of lowering the 
posted speed limit could refer only to the consequence of legislatures putting words on a page and have no meaningful impact on fatalities in and of itself. 
Drivers would have to be educated about the law, or the law would have to include provisions on redesigns and enforcement. If taken mechanistically, 
the effect of lowering the posted speed limit could refer to the physical state of the road. Vehicles traveling at a lower speed may be able to better maneuver 
and collide with less force. The effect then pertains to vehicles that comply perfectly with the reduced limit. In this paper, the effect of a countermeasure, 
such as lowering the posted speed limit, referred to the policy change and all subsequent actions to educate drivers and enforce it with road redesigns, 
traffic tickets and other countermeasures. 

Policymakers should consider multiple effects of Vision Zero policy. In theory, Vision Zero differs substantially from the traditional approach to traffic safety policy, 
and future work may seek to understand how drivers change their behavior as a result of specific policies. For example, traditional traffic safety 
improvements often widen roads and crosswalks to aid evasive maneuvers that reduce collisions, while Vision Zero advocates improvements that narrow 
them, much like the road ``diets'' of Goodwin et al. (2010), to force drivers to decrease their vehicle speed. Consequently, reducing fatalities under the 
Vision Zero approach may actually increase the number of collisions. For details see Johansson (2009).  

In considering these additional effects, a variety of methodological hurdles will come to light. Selection bias is just one of a litany of statistical issues that 
commonly arise in the estimation of causal effects, along with interference, temporal confounding and collision bias. Readers are referred to Hauer (2005) 
and Davis (2000) for a general discussion of the methodological challenges that arise in the evaluation of traffic safety policy.

This paper makes no policy recommendations. Triage may be the most sensible way to implement a Vision Zero strategy, but any subsequent 
evaluation must be performed with care. The U.S. Department of Transportation directs its offices to assign human life a value of roughly ten 
million dollars when conducting analysis that carries implications for public safety (Transportation 2016). The investment of billions of dollars 
over the lifetime of a strategy would then be considered effective if it were to save hundreds of lives. When making these and similar comparisons, 
it is important for policymakers to understand how to identify and adjust for selection bias. Failure can make an ineffective policy appear effective 
when perhaps a different strategy could prevent a greater number of fatalities or at a significantly reduced cost.

\subsection{6. Acknowledgement}\label{acknowledgement}

We would like to thank Roya Amjadi, Wendy Martinez, Stas Kolenikov and
the American Statistical Association's Government Statistics Section for
their encouragement. We would also like to thank Michael Sobel, Owen
Ward and members of New York City Community Board 7, especially Richard
Robbins and Catherine DeLazzero for their knowledge and expertise.

\subsection*{7. References}\label{references}
\addcontentsline{toc}{subsection}{8. References}

\hypertarget{refs}{}
\hypertarget{ref-aashto2001policy}{}
Aashto, A. 2001. ``Policy on Geometric Design of Highways and Streets.''
\emph{American Association of State Highway and Transportation
Officials, Washington, DC} 1 (990): 158.

\hypertarget{ref-trafficfacts2016}{}
Administration, National Highway Traffic Safety. 2016. ``Traffic Safety
Facts: Research Note, Dot Hs 812 260.''

\hypertarget{ref-berger2013statistical}{}
Berger, James O. 2013. \emph{Statistical Decision Theory and Bayesian
Analysis}. Springer Science \& Business Media.

\hypertarget{ref-davis2000accident}{}
Davis, Gary A. 2000. ``Accident Reduction Factors and Causal Inference
in Traffic Safety Studies: A Review.'' \emph{Accident Analysis \&
Prevention} 32 (1). Elsevier: 95--109.

\hypertarget{ref-efron2016computer}{}
Efron, Bradley, and Trevor Hastie. 2016. \emph{Computer Age Statistical
Inference}. Vol. 5. Cambridge University Press.

\hypertarget{ref-freedman1998statistics}{}
Freedman, David, Robert Pisani, and Roger Purves. 1998.
\emph{Statistics}. Vol. 4. WW Norton \& Company.

\hypertarget{ref-friedman1992old}{}
Friedman, Milton. 1992. ``Do old fallacies ever die?''.
\emph{Journal of Economic Literature}. Vol. 30, No.  4
American Economic Association: 2129--2132.

\hypertarget{ref-gelman2005analysis}{}
Gelman, Andrew. 2005. ``Analysis of Variance---why It Is More Important
Than Ever.'' \emph{The Annals of Statistics} 33 (1). Institute of
Mathematical Statistics: 1--53.

\hypertarget{ref-gelman2006data}{}
Gelman, Andrew, and Jennifer Hill. 2006. \emph{Data Analysis Using
Regression and Multilevel/Hierarchical Models}. Cambridge University
Press.

\hypertarget{ref-gelman1997poststratification}{}
Gelman, Andrew, and Thomas C Little. 1997. ``Poststratification into
Many Categories Using Hierarchical Logistic Regression.'' \emph{Survey
Methodology} 23 (2): 127--35.

\hypertarget{ref-good1953population}{}
Good, Irving J. 1953. ``The Population Frequencies of Species and the
Estimation of Population Parameters.'' \emph{Biometrika}. JSTOR,
237--64.

\hypertarget{ref-goodwin2010countermeasures}{}
Goodwin, Arthur H, Libby J Thomas, William L Hall, and Mary Ellen
Tucker. 2010. ``Countermeasures That Work: A Highway Safety
Countermeasure Guide for State Highway Safety Offices.''

\hypertarget{ref-greenwood1920inquiry}{}
Greenwood, Major and Yule, G Udny. 1920. ``An inquiry into the nature 
of frequency distributions representative of multiple happenings with 
particular reference to the occurrence of multiple attacks of disease or of 
repeated accidents'' \emph{Journal of the Royal statistical society} 83 (2): 255--79.

\hypertarget{ref-SwedGov2016}{}
Government Offices of Sweden, and The Swedish Trade \& Investment
Council. n.d. ``Vision Zero.''
\url{http://www.visionzeroinitiative.com/}.

\hypertarget{ref-hauer2005cause}{}
Hauer, Ezra. 2005. ``Cause and Effect in Observational Cross-Section
Studies on Road Safety.'' \emph{Unpublished Manuscript}.

\hypertarget{ref-imbens2015causal}{}
Imbens, Guido W, and Donald B Rubin. 2015. \emph{Causal Inference in
Statistics, Social, and Biomedical Sciences}. Cambridge University
Press.

\hypertarget{ref-johansson2009vision}{}
Johansson, Roger. 2009. ``Vision Zero--Implementing a Policy for Traffic
Safety.'' \emph{Safety Science} 47 (6). Elsevier: 826--31.

\hypertarget{ref-leaf1999literature}{}
Leaf, William A, and David F Preusser. 1999. \emph{Literature Review on
Vehicle Travel Speeds and Pedestrian Injuries}. US Department of
Transportation, National Highway Traffic Safety Administration.

\hypertarget{ref-mokdad2004actual}{}
Mokdad, Ali H, James S Marks, Donna F Stroup, and Julie L Gerberding.
2004. ``Actual Causes of Death in the United States, 2000.'' \emph{JAMA}
291 (10). American Medical Association: 1238--45.

\hypertarget{ref-national1998managing}{}
National Research Council (US). Transportation Research Board, Committee
for Guidance on Setting, and Enforcing Speed Limits. 1998.
\emph{Managing Speed: Review of Current Practice for Setting and
Enforcing Speed Limits}. Vol. 254. Transportation Research Board.

\hypertarget{ref-robbins1955empirical}{}
Robbins, Herbert. 1955. ``An Empirical Bayes Approach to Statistics.''
\emph{Proceedings of Third Berkeley Symp. Math. Statist. Probab.} 1 (1).
University of California Press, Berkeley: 157--64.

\hypertarget{ref-robbins1988estimating}{}
Robbins, Herbert, and Cun-Hui Zhang. 1988. ``Estimating a Treatment
Effect Under Biased Sampling.'' \emph{Proceedings of the National
Academy of Sciences} 85 (11). National Acad Sciences: 3670--2.

\hypertarget{ref-robbins2000efficiency}{}
---------. 2000. ``Efficiency of the U, V Method of Estimation.''
\emph{Proceedings of the National Academy of Sciences} 97 (24). National
Acad Sciences: 12976--9.

\hypertarget{ref-rosen2009pedestrian}{}
Rosén, Erik, and Ulrich Sander. 2009. ``Pedestrian Fatality Risk as a
Function of Car Impact Speed.'' \emph{Accident Analysis \& Prevention}
41 (3). Elsevier: 536--42.

\hypertarget{rosenbaum2017observation}{}
Rosenbaum, Paul R. 2017. ``Observation and experiment: an introduction 
to causal inference'' \emph{Harvard University Press} 2017.

\hypertarget{ref-RStan}{}
Stan Development Team. 2016. \emph{RStan: The R Interface to Stan}
(version 2.14.1). \url{http://mc-stan.org}.

\hypertarget{ref-stigler2016seven}{}
Stigler, Stephen M. 2016. \emph{The Seven Pillars of Statistical
Wisdom}. Harvard University Press.

\hypertarget{si2015bayesian}{}
Si, Yajuan and Natesh S Pillai and Andrew Gelman. 2015. ``Bayesian 
Nonparametric Weighted Sampling Inference.'' \emph{Bayesian Analysis}
10 (3). International Society for Bayesian Analysis: 605--625.

\hypertarget{ref-shelton1991national}{}
Shelton, Terry ST. 1991. ``National Accident Sampling System General Estimates 
System Technical Note, 1988 to 1990.''

\hypertarget{ref-vzthree}{}
Taskforce, New York City Vision Zero. 2017. ``Vision Zero: Year Three
Report.''
\url{http://www1.nyc.gov/assets/visionzero/downloads/pdf/vision-zero-year-3-report.pdf}.

\hypertarget{ref-tingvall2000vision}{}
Tingvall, Claes, and Narelle Haworth. 2000. ``Vision Zero: An Ethical
Approach to Safety and Mobility.'' In \emph{6th Ite International
Conference Road Safety \& Traffic Enforcement: Beyond}. Vol. 1999.

\hypertarget{ref-lifevalue2016}{}
Transportation, U.S. Department of. 2016. ``Revised Departmental
Guidance on Valuation of a Statistical Life in Economic Analysis.''
\url{http://www.transportation.gov/office-policy/transportation-policy/revised-departmental-guidance-on-valuation-of-a-statistical-life-in-economic-analysis/}.

\hypertarget{ref-xu2010national}{}
Xu, Jiaquan, Sherry L Murphy, Kenneth D Kochanek, and Brigham A.
Bastian. 2016. ``National Vital Statistics Reports.'' \emph{National
Vital Statistics Reports} 64 (2).

\subsection{8. Appendix}\label{appendix}

The following Stan Code was used in Subsection 4.2 (Stan Development Team 2016).

\begin{verbatim}
functions {
  int poisson_log_trunc_rng(real log_rate) {
    int draw;
    draw = poisson_log_rng(log_rate);
    while (draw == 0)
      draw = poisson_log_rng(log_rate);
    return draw;
  }
}
data {
  int<lower=1> N_train;
  int<lower=1> N;      
  int<lower=1> G;      
  int<lower=3> J[G];   
  int ints[N,G];       
  vector[N] EXPR;      
  int count[N];        
  int<lower=1> n_main;
  int<lower=1> n_inter;
}
transformed data {
  vector[N] offset = log(EXPR);
  int count_train[N_train] = count[1:N_train];
  int n_vars = G + 1;
}
parameters {
  real offset_e;
  vector[J[1]] eta_1;
  vector[J[2] - 1] eta_2;
  real nyc_e;
  vector[J[3]] eta_3;
  vector[J[4]] eta_4;
  vector[J[5]] eta_5;
  vector[J[6]] eta_6;
  vector[J[7]] eta_7;
  vector[J[8]] eta_8;
  vector[J[9]] eta_9;
  vector[J[10]] eta_10;
  vector[J[11]] eta_11;
  vector[J[12]] eta_12;
  vector[J[13]] eta_13;
  vector[J[14]] eta_14;
  vector[J[15]] eta_15;
  vector[J[16]] eta_16;
  vector[J[17]] eta_17;
  vector[J[18]] eta_18;
  vector[J[19]] eta_19;
  vector[J[20]] eta_20;
  vector[J[21]] eta_21;
  vector[J[22]] eta_22;
  vector[J[23]] eta_23;
  vector[J[24]] eta_24;
  vector[J[25]] eta_25;
  vector[J[26]] eta_26;
  vector[J[27]] eta_27;
  vector[J[28]] eta_28;
  vector[J[29]] eta_29;
  vector[J[30]] eta_30;
  vector[J[31]] eta_31;
  vector[J[32]] eta_32;
  vector[J[33]] eta_33;
  vector[J[34]] eta_34;
  vector[J[35]] eta_35;
  vector[J[36]] eta_36;
  vector[N_train] cell_eta;
  vector[n_main] eta_main;
  vector[n_inter] eta_inter;
  real<lower=0> sigma_sigma_main;
  real<lower=0> sigma_sigma_inter;
  real<lower=0> sigma_cell;
  real tau_main;
  real tau_inter;
  real mu;
}
transformed parameters {
  vector[J[1]] e_1;
  vector[J[2]] e_2;
  vector[J[3]] e_3;
  vector[J[4]] e_4;
  vector[J[5]] e_5;
  vector[J[6]] e_6;
  vector[J[7]] e_7;
  vector[J[8]] e_8;
  vector[J[9]] e_9;
  vector[J[10]] e_10;
  vector[J[11]] e_11;
  vector[J[12]] e_12;
  vector[J[13]] e_13;
  vector[J[14]] e_14;
  vector[J[15]] e_15;
  vector[J[16]] e_16;
  vector[J[17]] e_17;
  vector[J[18]] e_18;
  vector[J[19]] e_19;
  vector[J[20]] e_20;
  vector[J[21]] e_21;
  vector[J[22]] e_22;
  vector[J[23]] e_23;
  vector[J[24]] e_24;
  vector[J[25]] e_25;
  vector[J[26]] e_26;
  vector[J[27]] e_27;
  vector[J[28]] e_28;
  vector[J[29]] e_29;
  vector[J[30]] e_30;
  vector[J[31]] e_31;
  vector[J[32]] e_32;
  vector[J[33]] e_33;
  vector[J[34]] e_34;
  vector[J[35]] e_35;
  vector[J[36]] e_36;
  vector[n_vars] sds;
  vector[N_train] cell_e;
  vector[N_train] mu_indiv;
  vector[n_inter] sigma_inter;
  vector[n_main] sigma_main;
  
  sigma_main = exp(-1 + 0.5 * tau_main + 0.3 * sigma_sigma_main * eta_main);
  sigma_inter = exp(-2 + 0.5 * tau_inter + 0.3 * sigma_sigma_inter * eta_inter);
  sds[1:n_main] = sigma_main;
  sds[(n_main + 1):(n_inter + n_main)] = sigma_inter;
  sds[n_vars] = 0.3 * sigma_cell;
  
  e_1 = sds[1] * eta_1;
  e_2[1:(J[2] - 1)] = sds[2] * eta_2;
  e_2[J[2]] = nyc_e;
  e_3 = sds[3] * eta_3;
  e_4 = sds[4] * eta_4;
  e_5 = sds[5] * eta_5;
  e_6 = sds[6] * eta_6;
  e_7 = sds[7] * eta_7;
  e_8 = sds[8] * eta_8;
  e_9 = sds[9] * eta_9;
  e_10 = sds[10] * eta_10;
  e_11 = sds[11] * eta_11;
  e_12 = sds[12] * eta_12;
  e_13 = sds[13] * eta_13;
  e_14 = sds[14] * eta_14;
  e_15 = sds[15] * eta_15;
  e_16 = sds[16] * eta_16;
  e_17 = sds[17] * eta_17;
  e_18 = sds[18] * eta_18;
  e_19 = sds[19] * eta_19;
  e_20 = sds[20] * eta_20;
  e_21 = sds[21] * eta_21;
  e_22 = sds[22] * eta_22;
  e_23 = sds[23] * eta_23;
  e_24 = sds[24] * eta_24;
  e_25 = sds[25] * eta_25;
  e_26 = sds[26] * eta_26;
  e_27 = sds[27] * eta_27;
  e_28 = sds[28] * eta_28;
  e_29 = sds[29] * eta_29;
  e_30 = sds[30] * eta_30;
  e_31 = sds[31] * eta_31;
  e_32 = sds[32] * eta_32;
  e_33 = sds[33] * eta_33;
  e_34 = sds[34] * eta_34;
  e_35 = sds[35] * eta_35;
  e_36 = sds[36] * eta_36;
  cell_e = sds[G+1] * cell_eta;

  for (n in 1:N_train)
    mu_indiv[n] = mu + offset_e * offset[n]
               + e_1[ints[n,1]]
               + e_2[ints[n,2]]
               + e_3[ints[n,3]]
               + e_4[ints[n,4]]
               + e_5[ints[n,5]]
               + e_6[ints[n,6]]
               + e_7[ints[n,7]]
               + e_8[ints[n,8]]
               + e_9[ints[n,9]]
               + e_10[ints[n,10]]
               + e_11[ints[n,11]]
               + e_12[ints[n,12]]
               + e_13[ints[n,13]]
               + e_14[ints[n,14]]
               + e_15[ints[n,15]]
               + e_16[ints[n,16]]
               + e_17[ints[n,17]]
               + e_18[ints[n,18]]
               + e_19[ints[n,19]]
               + e_20[ints[n,20]]
               + e_21[ints[n,21]]
               + e_22[ints[n,22]]
               + e_23[ints[n,23]]
               + e_24[ints[n,24]]
               + e_25[ints[n,25]]
               + e_26[ints[n,26]]
               + e_27[ints[n,27]]
               + e_28[ints[n,28]]
               + e_29[ints[n,29]]
               + e_30[ints[n,30]]
               + e_31[ints[n,31]]
               + e_32[ints[n,32]]
               + e_33[ints[n,33]]
               + e_34[ints[n,34]]
               + e_35[ints[n,35]]
               + e_36[ints[n,36]]
               + cell_e[n];
}
model {
  eta_1 ~ normal(0, 1);
  eta_2 ~ normal(0, 1);
  eta_3 ~ normal(0, 1);
  eta_4 ~ normal(0, 1);
  eta_5 ~ normal(0, 1);
  eta_6 ~ normal(0, 1);
  eta_7 ~ normal(0, 1);
  eta_8 ~ normal(0, 1);
  eta_9 ~ normal(0, 1);
  eta_10 ~ normal(0, 1);
  eta_11 ~ normal(0, 1);
  eta_12 ~ normal(0, 1);
  eta_13 ~ normal(0, 1);
  eta_14 ~ normal(0, 1);
  eta_15 ~ normal(0, 1);
  eta_16 ~ normal(0, 1);
  eta_17 ~ normal(0, 1);
  eta_18 ~ normal(0, 1);
  eta_19 ~ normal(0, 1);
  eta_20 ~ normal(0, 1);
  eta_21 ~ normal(0, 1);
  eta_22 ~ normal(0, 1);
  eta_23 ~ normal(0, 1);
  eta_24 ~ normal(0, 1);
  eta_25 ~ normal(0, 1);
  eta_26 ~ normal(0, 1);
  eta_27 ~ normal(0, 1);
  eta_28 ~ normal(0, 1);
  eta_29 ~ normal(0, 1);
  eta_30 ~ normal(0, 1);
  eta_31 ~ normal(0, 1);
  eta_32 ~ normal(0, 1);
  eta_33 ~ normal(0, 1);
  eta_34 ~ normal(0, 1);
  eta_35 ~ normal(0, 1);
  eta_36 ~ normal(0, 1);
  cell_eta ~ normal(0, 1);
  offset_e ~ normal(0, 1);
  mu ~ normal(-10,3);
  nyc_e ~ normal(0, 2);
  
  sigma_sigma_inter ~ normal(0, 1);
  sigma_sigma_main ~ normal(0, 1);
  
  eta_inter ~ normal(0, 1);
  eta_main ~ normal(0, 1);
  
  tau_inter ~ normal(0, 1);
  tau_main ~ normal(0, 1);
  
  sigma_cell ~ normal(0, 1);
  
  for (n in 1:N_train)
    target += -log1m_exp(-exp(mu_indiv[n])) + poisson_log_lpmf(count_train[n] | mu_indiv[n]);
}
generated quantities {
  real sd_1;
  real sd_2;
  real sd_3;
  real sd_4;
  real sd_5;
  real sd_6;
  real sd_7;
  real sd_8;
  real sd_9;
  real sd_10;
  real sd_11;
  real sd_12;
  real sd_13;
  real sd_14;
  real sd_15;
  real sd_16;
  real sd_17;
\end{verbatim}

\end{document}